\documentclass[acmsmall, nonacm]{acmart}

\usepackage{caption}
\usepackage{subcaption}
\usepackage{todonotes}
\usepackage{pifont}
\usepackage{xspace}
\usepackage{CJKutf8}
\usepackage{makecell}
\usepackage{enumitem}
\usepackage{listings}
\usepackage{amsmath}
\usepackage[ruled,linesnumbered]{algorithm2e}
\newcolumntype{L}[1]{>{\raggedright\let\newline\\\arraybackslash\hspace{0pt}}m{#1}}
\newcolumntype{C}[1]{>{\centering\let\newline\\\arraybackslash\hspace{0pt}}m{#1}}
\newcolumntype{R}[1]{>{\raggedleft\let\newline\\\arraybackslash\hspace{0pt}}m{#1}}
\newcommand{\zh}[1]{\begin{CJK}{UTF8}{gbsn}#1\end{CJK}}

\newcommand\promptbox[1]{\todo[inline,color=gray!10]{#1}}
\newcommand\systemmessage{\noindent \textbf{[SYSTEM MESSAGE]:}}
\newcommand\usermessage{\noindent \textbf{[USER MESSAGE]:}}

\newcommand{\ie}{{\em i.e.,\/ }}
\newcommand{\eg}{{\em e.g.,\/ }}
\newcommand{\vs}{{\em vs.\/ }}
\newcommand{\etc}{{\em etc. \/}}

\newcommand{\pb}[1]{\vspace{0.75ex}\noindent{\bf \em #1}\hspace*{.3em}}
\newcommand\blue[1]{{#1}}

\newcommand\minorblue[1]{{#1}}

\newcommand{\one}{({\em i}\/)\xspace}
\newcommand{\two}{({\em ii}\/)\xspace}
\newcommand{\three}{({\em iii}\/)\xspace}
\newcommand{\four}{({\em iv}\/)\xspace}
\newcommand{\five}{({\em v}\/)\xspace}

\AtBeginDocument{%
  }

\begin{document}

\title{Understanding Fanchuan in Livestreaming Platforms: A New Form of Online Antisocial Behavior}


\author{Yiluo Wei}
\authornote{Joint First Authors}
\affiliation{%
  \institution{The Hong Kong University of Science and Technology
(Guangzhou)}
  \country{China}}

\author{Jiahui He}
\authornotemark[1]
\affiliation{%
  \institution{The Hong Kong University of Science and Technology
(Guangzhou)}
  \country{China}}

\author{Gareth Tyson}
\affiliation{%
  \institution{The Hong Kong University of Science and Technology
(Guangzhou)}
  \country{China}}







\renewcommand{\shortauthors}{Yiluo Wei, Jiahui He, and Gareth Tyson}

\begin{abstract}
Recently, a distinct form of online antisocial behavior, known as ``fanchuan'', has emerged across online platforms, particularly in livestreaming chats. Fanchuan is an indirect attack on a specific entity, such as a celebrity, video game, or brand. It entails two main actions: 
\one~individuals first feign support for the entity, and exhibit this allegiance widely;
\two~they then engage in offensive or irritating behavior, attempting to undermine the entity by association.
This deceptive conduct is designed to tarnish the reputation of the target and/or its fan community.
Fanchuan is a novel, covert and indirect form of social attack, occurring outside the targeted community (often in a similar or broader community), with strategic long-term objectives. This distinguishes fanchuan from other types of antisocial behavior and presents significant new challenges in moderation. 
We argue it is crucial to understand and combat this new malicious behavior. Therefore, we conduct the first empirical study on fanchuan behavior in livestreaming chats, focusing on Bilibili, a leading livestreaming platform in China. Our dataset covers 2.7 million livestreaming sessions on Bilibili, featuring 3.6 billion chat messages. We identify 130k instances of fanchuan behavior across 37.4k livestreaming sessions. Through various types of analysis, our research offers valuable insights into fanchuan behavior and its perpetrators.
\end{abstract}






\maketitle

\section{Introduction}

Live streaming is gaining significant popularity globally, encompassing a wide range of topics including personal experiences \cite{lottridge2017thirdwave, lu2019vicariously, tang2016meerkat}, artistic creation \cite{fraser2019sharing, lu2019responsibility}, educational content \cite{lu2018watch, faas2018watch, lu2018streamwiki}, and gaming \cite{hamilton2014streaming, 7382994}. 
Among the many features enhancing the viewer experience in livestreaming, bullet chat (\ie danmaku) \cite{ZHOU2019100815, ZHANG2024103793} has become particularly popular, where chat messages from the audience are presented together on the video feed and automatically scroll in real-time. This creates an energetic, communal atmosphere that draws in participants and fosters a sense of shared experience \cite{10.1145/2556288.2557048, 10.1145/3027063.3052765, 9604216, WANG2024101397}. 

However, the very features that make livestreaming so engaging also open the door to negative behaviors. Toxicity in chat messages is a growing issue, where inappropriate language, harassment, and disruptive comments can quickly escalate, affecting both the streamer and the audience \cite{10.1145/3610191, 10.1145/3452918.3458794, 10333159, Jiang_Shen_Wen_Sha_Chu_Liu_Backes_Zhang_2024}.  
The need for effective moderation is clear, but managing the volume and variety of messages in a livestream is a formidable challenge. The real-time nature of the chats also make moderation errors irreversible. Thus, automated moderation is a formidable task \cite{moon-etal-2023-analyzing, Jiang_Shen_Wen_Sha_Chu_Liu_Backes_Zhang_2024}, often making it necessary for manual moderation by humans. Moderators must balance the need to maintain a positive environment without stifling genuine conversation.
Consequently, moderation is a crucial yet complex aspect of the livestreaming experience \cite{10.1145/3452918.3458796, 10.1145/3491102.3517628, 10.1145/3567568, 10.1145/3479554, 10.1145/3544549.3585704}. 

Recently, a specific form of malicious behavior, known as \emph{``fanchuan''} (\zh{反串} in Chinese), has emerged on various platforms, especially in livestreaming chats.
Fanchuan is an indirect attack on a specific entity (\eg a ``celebrity'' streamer, a video game, or a brand), which involves two main actions: 
\one Individuals \emph{pretend} to support the entity and display this support to others; 
then
\two They engage in actions that are \emph{offensive or irritating}, in an attempt to damage the reputation of the target entity and/or its fan community.
For example, in the case of a video game $G$, a fanchuan attack might involve someone sending a message during another video game livestream: ``We players of $G$ [pretending] do not play this trash game [offensive and irritating]''.

Unlike other forms of antisocial behavior \cite{10.1145/3555552}, such as trolling \cite{10.1145/2998181.2998213}, flaming \cite{Jane2015}, hate \cite{WALTHER2022101298}, and harassment \cite{10.1145/3134659}, fanchuan is a \emph{covert} and \emph{indirect} attack. It occurs \emph{outside} the targeted community (usually in a similar or broader community) and strategically aims for \emph{long-term} effects. 
In contrast, other antisocial behaviors are typically \emph{overt} and \emph{direct}, targeting individuals \emph{within} the community and often seeking \emph{immediate} harm, sometimes at random or for fun \cite{ShachafHara2010, VarjasEtAl2010}. That said, fanchuan can also incorporate other forms of antisocial behavior in their offensive performance, although this is merely the approach, not their true intention.

Fanchuan presents a significant challenge in moderation. First, like other antisocial behaviors, it can harm livestreams by creating a deceptive and toxic environment. The key difference, however, is that even if moderation occurs, unless it happens preemptively, the attack can still succeed. Other viewers of the livestream are likely to recognize that ``a fan of XX was acting foolishly/trolling/flaming  and got blocked'', which means that the goal of damaging the reputation of the entity and/or its fans is still reached.
Therefore, in addition to blocking, it is crucial to accurately identify and inform others that this is actually a fanchuan behavior, to effectively counteract its negative impact. However, a fanchuan chat message often exists in a gray area,  making it difficult to identify whether it is from a genuine (but somewhat exaggerated) fan, or a fanchuan user that pretends to be a fan, especially for traditional automated moderation tools. Even human moderators need background knowledge about the target to discern the true intent. Since fanchuan behavior typically occurs outside the targeted community, human moderators may also struggle to recognize it.

We therefore argue that it is crucial to understand and combat the detrimental effects of this unique type of malicious behavior. To the best of our knowledge, no prior works have studied this new form of abuse. To bridge this gap, we conduct the first empirical study on fanchuan behavior in livestreaming chat, with a focus on the Bilibili platform. Bilibili is a prominent live streaming platform in China, and is the first site in the country to offer the innovative bullet chat feature. Our dataset comprises 2.7 million live streaming sessions on Bilibili with 3.6 billion chat messages, covering the period from June 2022 to August 2023. These sessions are from the \minorblue{most popular streamers on Bilibili (\ie those with a large fan base, usually more than 50k followers)}. We identify 130k fanchuan attacks occurring in 37.4k livestreaming sessions. With this dataset as our foundation, we explore the following research questions:

\begin{itemize}[leftmargin=*]

    \item \textbf{RQ1}:  What are the types of livestreams in which fanchuan typically occurs, and what are the subjects commonly associated with fanchuan?

    \item \textbf{RQ2}:
    \blue{What are the immediate impacts of fanchuan on livestreaming session chats, including alterations in the quantity, sentiment, and toxicity of the messages?}

    \item \textbf{RQ3}:
    \blue{Do users who engage in fanchuan exhibit distinct activity and chat patterns when compared to other users on Bilibili?}

    \item \textbf{RQ4}: 
    \blue{Is it possible to train a machine learning model that can automatically identify users who are likely to engage in fanchuan?}

\end{itemize}

Through studying these RQs, our \blue{contributions} include:

\begin{enumerate}[leftmargin=*]
    \item 
    \blue{We show the complexity and potential widespread impact of fanchuan, emphasizing the need for better understanding and moderation:} 
    The targets of fanchuan attacks span a diverse range of areas and online communities. While fanchuan attacks focus on stream types known for conflict, such as esports. It also extends to areas that might initially seem less susceptible to such behaviors, like Gacha \cite{10.1145/3579438} games. The targets range from well-known (e)sports players to virtual anime characters. (\S\ref{sec:RQ1})

    \item 
    \blue{We show the immediate (harmful) impact driven by fanchuan chat messages:}
    There is a significant increase in the quantity of chat messages after the fanchuan behavior, but the impact does not last for a long period (around 2 minutes). During this surge, there is a large amount of duplicated chat messages, and an increase in messages with negative sentiment or toxicity.  (\S\ref{sec:RQ2})

    \item 
    \blue{We show the distinct characteristics of users who engage in fanchuan behavior:}
    They tend to do so repeatedly, with 88\% exhibiting this repeat pattern, often targeting one specific entity (94\%); and the majority of fanchuan cases (70\%) occur more than five minutes after the user begins watching the livestream, suggesting these users are more akin to ``ordinary viewers'' rather than random spammers. 
    Additionally, they exhibit lower levels of platform-wide activity compared to other users, which is primarily because their activity is concentrated on the livestreams of one specific theme (the area where they engage in Fanchuan behavior). Moreover, they show a higher levels of toxicity and negativity in their chat messages compared to other users. (\S\ref{sec:RQ3})

    \item Based on the above findings, we demonstrate that \blue{our machine learning model can effectively identify} a small subset (\eg 50) of users likely to send fanchuan messages out of hundreds or thousands of viewers. \blue{This capability makes it feasible to develop an automated tool for moderators, allowing them to streamline the manual review of these users' messages and block potential fanchuan content.} (\S\ref{sec:rq4})
    
\end{enumerate}

\section{Background \& Related Work}
\label{sec:background}

\subsection{Fanchuan \& Other Antisocial Behavior}

Online antisocial behavior can be seen as a continuation of analogous actions that occur offline. This encompasses various kinds of aggressive acts, harassment, and bullying \cite{AkbulutEtAl2010, Kayany1998}. The reasons behind this behavior, along with the reactions to it, have also been researched and reviewed \cite{KieslerEtAl2012, 9519435}. In this subsection, we introduce fanchuan, a distinctive and newly emerging form of online antisocial behavior, and compare it with other related types of antisocial behavior.

\pb{Defining Fanchuan.}
Fanchuan is an indirect attack on a specific entity, such as a celebrity, a video game, or a brand. It involves two main actions: \one Individuals feign support for the entity and publicly display this support. \two They engage in behaviors that are offensive or irritating. This deceptive behavior aims to ruin the reputation of the entity and its fan community by casting them in a negative light through their own actions.

Fanchuan attackers typically publicly demonstrate their feigned support for an entity via the content of their posted messages. This approach showcases their (fake) support either explicitly or implicitly, along with irritating content. For instance, a fanchuan attack on a game, $G$, might involve statements like, ``We players of $G$' will never play your trash game'' (explicit), or ``Why not play $G$? It's a much better game'' (implicit). In these examples, regular users who encounter these messages might perceive the fans of ($G$) as arrogant and offensive, thereby damaging the reputation of both the game and its fan community.

\pb{Comparing to Toxic Content.}
According to the review by Thomas et. al. \cite{9519435}, toxic content attacks can undermine availability by preventing victims from effectively participating in an online community, potentially even driving them away. These attacks encompass a wide range of antisocial behaviors, such as trolling \cite{10.1145/2998181.2998213}, flaming \cite{Jane2015}, hate speech \cite{WALTHER2022101298}, and harassment \cite{10.1145/3134659}. A shared characteristic of these attacks is that they have a specific target, typically an individual user (may also be a community), and are intended to be seen by the target. The toxic content is either directed at the target user or posted within the target community.

This is the key difference from fanchuan attacks, where the target is not a typical user but a broader concept, which could be a person, game, company, and brand, together with their fan community. Additionally, unlike toxic content attacks, fanchuan attacks are not meant to be seen by the target but are directed at users in a different community. However, it is important to note that fanchuan attacks can also incorporate toxic content as a method to provoke or annoy others. Consequently, the negative impact of toxic content attacks can manifest in fanchuan attacks as a less intended ``side effect''.

\pb{Comparing to Coordinated Attack.}
Some attacks inherently require coordinated action or amplification to succeed. Notable examples include organized trolling activities orchestrated on platforms like Facebook \cite{phillips2011loling}, Reddit \cite{kumar2018community}, and 4chan \cite{hine2017kek}. 
Other examples are raids, where a large group overwhelms the comment section of a targeted group or individual \cite{10.1145/3359309, 10.1145/3610191}, and dogpiling, where a person is pressured to retract an opinion or statement \cite{Kang2022Closing}.

These are rather different from fanchuan, as these approaches typically attack the targets directly, whereas fanchuan attacks are far more subtle and indirect. 
That said, the similarity is that fanchuan attacks also require numerous actions to reach a wide audience and effectively damage the target's reputation. This necessitates the involvement of many fanchuan attackers. However, fanchuan attacks require significantly less, or even no coordination among the attackers. This is mainly because fanchuan is a long-term, distributed attack, which eliminates the need for precise coordination to strike a single point at a specific moment.

\pb{Comparing to Impersonation.}
Impersonation occurs when an attacker deceives an audience by adopting the online persona of a target to create content that damages the target's reputation or causes emotional harm \cite{10.1145/3025453.3025875, 10.1145/3173574.3174241}.
This concept is similar to fanchuan, but impersonation involves pretending to be a specific person, \ie the target, while fanchuan attacks do not involve impersonating a specific individual. Consequently, fanchuan attacks require significantly less preparation, such as setting up an dedicated account or even stealing one from the target.
Despite this, they both share the characteristic of not directly attacking the target. Intuitively, in some cases, fanchuan could be seen as an ``impersonation of a community'', where the attacker deceives the audience into believing they represent the target's fan community.

\subsection{Content Moderation in Livestreaming}

The real-time nature of livestreaming chat makes it difficult to moderate. In response, prior research has attempted to study content moderation for livestreaming. 

\pb{Manual Review Solutions.}
Uttarapong \emph{et al.} \cite{10.1145/3452918.3458794} examined the harassment experiences and response strategies of marginalized (women and LGBTQ+) streamers on Twitch, highlighting that marginalized streamers often depend on human effort, and platforms lacking adequate technical support to handle harassment issues.
Thach \emph{et al.} \cite{thach2024visible} explored content moderation practices on Reddit and Twitch, focusing on how visibility impacts marginalized users. On Twitch, moderation methods include human real-time oversight during livestreaming chats and interactions from streamers, contrasting with Reddit's reliance on volunteer moderators and automated tools that often make moderation invisible to users.
However, these manual review methods may not be effective at identifying fanchuan behavior. The massive volume of chat message generated by livestreaming users and the complex nature of fanchuan makes manual review infeasible. It also requires that the moderator has sufficient background knowledge to accurately identify fanchuan behavior.

\pb{Machine Learning Solutions.}
Tarafder \emph{et al.} \cite{tarafder2023automated} proposed an automated moderation tool to filter toxic comments in livestreaming chats on platforms like YouTube. The tool uses the YouTube API to help moderators maintain a healthy environment, achieving 97\% accuracy in English, with plans for multilingual support.
Moon \emph{et al.} \cite{moon-etal-2023-analyzing} proposed an NLP method for detecting norm violations in livestreaming chats. The study shows that some additional information, such as the context of chats and videos, is a key feature for detecting norm violations, and that properly contextualized information can improve detection performance by up to 35\%.
Tang \emph{et al.} \cite{tang2021videomoderator} proposed VideoModerator, which is a risk-aware framework designed to enhance the moderation of e-commerce livestreaming videos by integrating human insights with machine learning, facilitating the identification of deviant content through interactive multi-modal visualizations. Its approach, including a ``learning with reviewing'' strategy and an intuitive interface, allows moderators to quickly navigate and assess video content effectively.
These show that machine learning is a valuable tool for detecting harmful behavior. However, this approach requires a comprehensive understanding of such harmful behavior, as well as a dedicated dataset to train the machine learning model. To the best of our knowledge, neither criterion has been met for moderating fanchuan behavior, and this is the first study to focus on fanchuan.

\subsection{Primer on the Bilibili Livestreaming Platform}

As a prominent streaming platform in China, Bilibili offers a wide variety of content creators including those focused on sports, esports, gaming, arts, Vtubers \cite{10.1145/3696410.3714803}, and more. Bilibili provides numerous livestreaming features, of which bullet chat, superchat, and user blocking which are related to our analysis.

\begin{figure}[h!]
    \centering
    \includegraphics[width=0.6\linewidth]{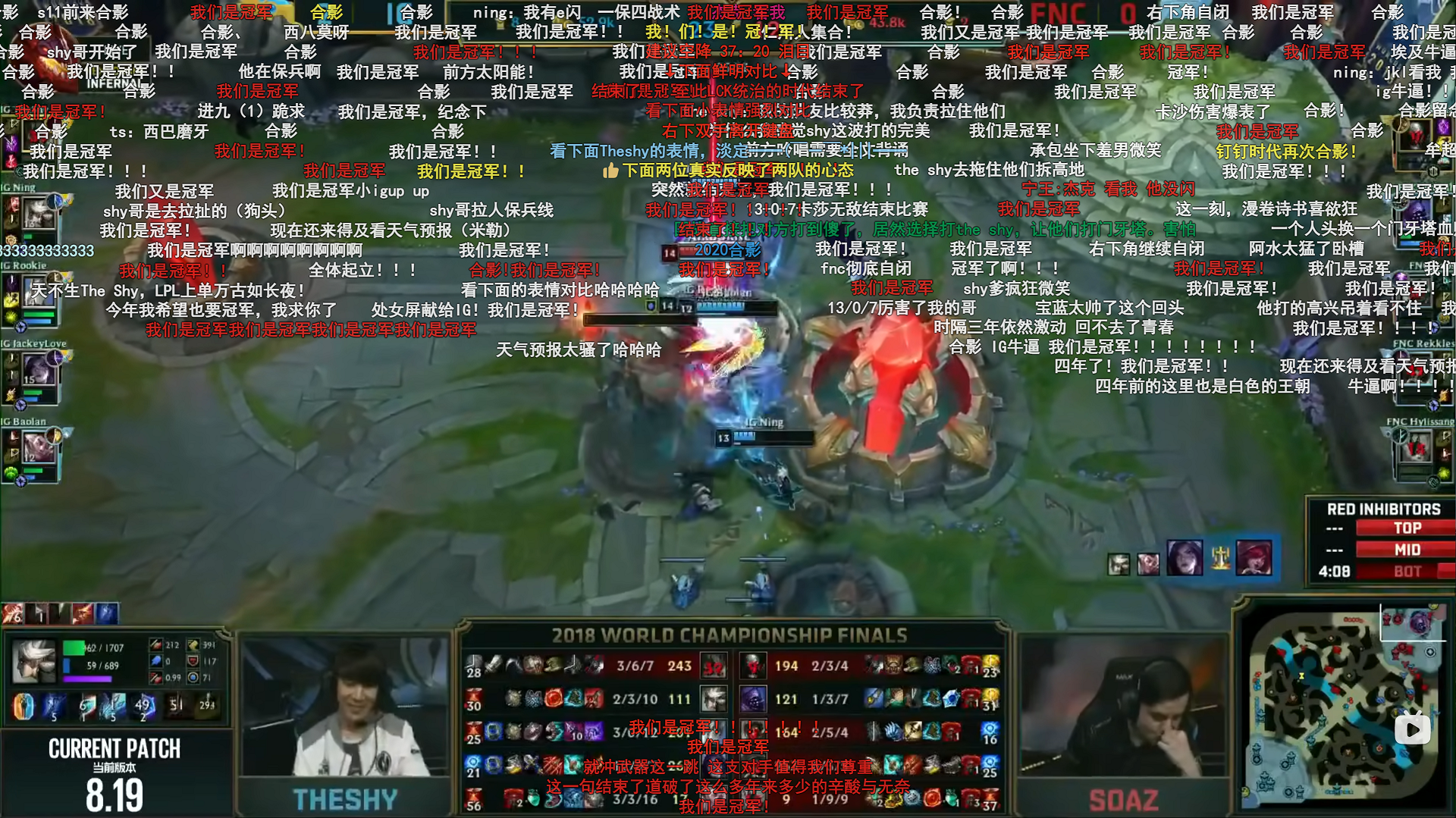}
    \caption{Screenshot of livestreaming  on Bilibili with bullet chats (the floating texts).}
    \label{fig:bullet_chat}
\end{figure}

\pb{Bullet Chat.}
Bullet chat, also known as danmaku in Japanese and danmu in Chinese, is a unique comment system that originated from the Japanese website Niconico. This interactive feature allows viewers to post comments directly onto the screen during a livestream, where they are displayed as moving text, as shown in Figure \ref{fig:bullet_chat}. Unlike traditional comment systems found on platforms like YouTube, bullet chat provides a more immersive and real-time engagement experience for users. As bullet chats are timely and direct, they often result in a far higher volume of comments from viewers during live streaming events \cite{huang2023}.
\blue{Previous research has examined viewer engagement with live streaming bullet chats, emphasizing their influence on virtual gift-sending \cite{LI2021101624, ZHOU2019100815}. Additionally, previous research has investigated the toxicity of bullet chats, underscoring the issue of toxic interactions during esports competition livestreams \cite{Jiang_Shen_Wen_Sha_Chu_Liu_Backes_Zhang_2024}. We also note that, although online videos and livestreaming might be two different scenarios, some research has explored bullet chats \cite{10.1145/3148330.3148344, 10.1145/3329485} and their moderation methods \cite{Hu2024DanModCapDA} in the context of online videos.}

\pb{Superchat.}
There is a Superchat (SC) system in place that allows users to pay for sending a special message that will be pinned at the top of the chat column for a specific period of time. This feature provides viewers with a way to ensure that both the streamer and other viewers read their message, and also gives streamers additional methods for generating revenue.

\pb{User Blocking.}
The streamer and the administrator of the live channel have the ability to block users, which will result in a message indicating that a user has been blocked.

\pb{Area \& Parent Area.}
The streamer must select from a list of parent areas and (child) areas to indicate the topic/theme of their livestream. For example, the streamer could select ``online game'' as the parent area and ``League of Legends'' as the area.

\section{Dataset \& Methodology}
\label{sec:dataset}

\subsection{Data Collection}
\label{subsec:data_collection}

\pb{Target Streamers.}
First, we compile a list of target streamers on Bilibili. To accomplish this, we rely on four streamer indexing and archiving sites: \texttt{VTBs.moe}, \texttt{laplace.live}, \texttt{zeroroku.com}, and \texttt{danmakus.com}.
We extract all streamers listed on the aforementioned sites and obtain a list of 26k streamers, which comprises the most popular streamers on Bilibili (\ie those with a relatively large \minorblue{fan base, usually with more than 50k followers}).

\pb{Livestreaming Sessions.}
For each selected streamer, we gather data on all their live streaming sessions between \texttt{2022-06-01} and \texttt{2023-09-01}. This data includes details such as session time, duration, title, category, \etc Additionally, our dataset includes viewer interactions during these livestreams, such as joining the stream, chatting, sending gifts, using super chat, subscribing for membership, and more. A full data description is available in the Appendix \ref{subsec:appendix_data_description}.
Overall, we have collected data for 2.7 million livestreaming sessions, with 10.7 billion interaction records and 3.6 billion chat messages.

\subsection{Dataset Construction}
\label{subsec:dataset_construction}

\pb{Identifying Fanchun Behavior by Keyword Search.}
To identify fanchuan behavior in our dataset, we rely on the observation that some other viewers often recognize this behavior and point this out in the chat. This is a form of ``crowd sourced'' moderation. 
This is typically done by publicly announcing that fanchuan has occurred, and that other users should ignore it.
Therefore, as the first step, we search for words related to fanchuan within all chat messages in our dataset.
We select three keywords: ``fanchuan'' (\zh{反串}), ``chuanzi'' (\zh{串子}), a derogatory term for individuals engaging in ``fanchuan'' behavior), and ``biechuan'' (\zh{别串}, stop ``fanchuan'').
In total, we identify 320k chat messages containing one of these keywords, indicating the potential occurrence of fanchuan behavior around the timestamp of these messages.

\pb{Result Filtering.}
It is important to note that the presence of words related to fanchuan does not necessarily guarantee that fanchuan behavior is occurred. For example, these words could be part of a general discussion about fanchuan.
To address this issue, we utilize an LLM to identify chat messages that do not suggest a fanchuan behavior happened in the livestreaming session. Specifically, we employ OpenAI's GPT-4o-mini model with a prompt outlined in Appendix \ref{subsec:appendix_prompts_filt}.
\blue{Human verification is then conducted to validate the \minorblue{LLM} predictions.
Specifically, the authors manually check 200 samples to assess the accuracy of the labels generated by the LLM. This evaluation demonstrates that the LLM achieves an F1-score of 95.5\%, indicating its effectiveness as a labeler. }

Note, a single fanchuan instance may be flagged by multiple users across several chat messages. Therefore, we merge chat messages that are close in time (within 30 seconds) and only keep the first one, as they likely refer to the same fanchuan instance. After merging, 130k fanchuan cases remain for our analysis.

\pb{Identifying Users Who Send the Fanchuan chat Message.}
After identifying the occurrence of fanchuan, we wish to determine which user was engaging in the behavior. This task is challenging due to the large volume of chats that appear around the timestamp where fanchuan occurs, many of which contain slang and abbreviations, making it difficult even for human labelers.

To ensure the reliability of our results, we employ two methods. 
First, fanchuan chats are sometimes sent using a superchat and are pointed out by others. In these cases, we only need to examine the superchats around the timestamp, typically finding only a small number of superchats (usually one). We then manually identify the fanchuan message for the cases where there are more than one superchats. Using this method, we identify 1,063 fanchuan chat messages.
Second, around the timestamp of a fanchuan incident (within 1 minute), some users may be blocked. User blocking is quite rare, with few or no instances in most livestreaming sessions. Thus, intuitively, we expect that these blocked users were engaging in fanchuan behavior. We identify 16,274 chat messages based on this criterion. In total, we identify 17,337 fanchuan chat messages sent by 5,267 users. Note, the chats do not contain direct references to specific users. Therefore, we cannot use @ mentions to identify the fanchuan user directly.

Based on the fanchuan messages, we finally compile two separate lists, a \one Fanchuan User List, and \two Comparison User List (of non-fanchuan users).
Specifically, for each identified fanchuan chat message, \one we add its sender to the Fanchuan User List, with the livestreaming session where the message is sent; 
and \two for the Comparison User List, we add in all remaining viewers (563 viewers on average) from the same livestreaming session who have sent at least one chat message within a 5-minute window preceding the fanchuan message.
Note, a user can appear multiple times if they participate in different livestreaming sessions.
For the analyses in the subsequent sections, users in the Fanchuan User List are referred to as ``\emph{\textbf{Fanchuan Users}}'', and users in the Comparison User List are refereed as ``\emph{\textbf{comparison Users}}''. Additionally, the Fanchuan Users who send the fanchuan message with superchat are referred as ``\emph{\textbf{Fanchuan (SC) Users}}''. Comparison Users who have sent a superchat during the measurement period are referred as ``\emph{\textbf{Comparison (SC) Users}}''.

\pb{Dataset Summary.}
Overall, our dataset consists of three parts. First, it includes 130k chat message records that indicate when fanchuan behavior occurred around its timestamp. Second, it contains 37,435 livestreaming session records where the (multiple) fanchuan behavior took place. Third, it includes the Fanchuan User List of 17,337 Fanchuan Users (5,267 unique) and the Comparison User list of 9.7 million users for comparison, along with all of their activity records during the measurement period (\texttt{2022-06-01} to \texttt{2023-09-01}).

\subsection{\blue{Considerations on Dataset Construction}}

\blue{We note that there may be multiple potential methdods to label our dataset. Here, we briefly discuss and justify our adopted method for dataset construction in \S\ref{subsec:dataset_construction}.} 

\pb{\blue{Why Not Human Labeling?}}
\blue{An obvious approach would be to perform human annotation of the data. However, due to the large amount (3.6 billion) of chat messages, it is impractical to follow this approach.}

\pb{\blue{Why Not Machine Learning?}}
\blue{First, it is important to note that there are currently no available models specifically designed for fanchuan detection or tasks that closely resemble it. Consequently, if we decide to employ machine learning methods, we will need to compile a relatively large labeled dataset for fanchuan, which necessitates a scalable method for detecting fanchuan. This situation presents a classic chicken-and-egg problem.
Moreover, previous studies have shown that machine learning approaches face some challenges when applied to the context of livestreaming chats \minorblue{(\eg previous studies indicate that widely used moderation services like the Perspective API and the OpenAI moderation API, perform poorly when applied to livestreaming chats \cite{moon-etal-2023-analyzing, Jiang_Shen_Wen_Sha_Chu_Liu_Backes_Zhang_2024}.)} and the context of Chinese Internet languages \minorblue{due to the highly flexible methods such as emoji-homophone replacement to cloak the content \cite{xiao-etal-2024-toxicloakcn}.} Therefore, even if we successfully compile a labeled dataset and train a machine learning model, its performance remains uncertain.
Therefore, we argue that machine learning methods is not a feasible approach for this task.}

\pb{\blue{Why Keyword Search?}}
\blue{First, we would like to re-emphasize that keyword searches are not employed to find fanchuan chat messages directly, but rather to locate messages that flag fanchuan behavior. In other words, this method is actually human labeling, as we are seeking ``crowd-sourced'' human labels (produced by other viewers) for fanchuan behavior. Moreover, these labels are likely created by the most appropriate labelers —-- the viewers —-- who possess the relevant knowledge and experience of the context. Therefore, we argue that this approach is currently the most suitable (and likely the only feasible) method for identifying fanchuan behavior on a large scale.}

\pb{\blue{Are there Sufficient Labelers in each Stream to Flag Fanchuan Behavior?}}
\blue{%
As described in \S\ref{subsec:data_collection}, our dataset consists of livestreams by streamers with relatively large fan bases. Consequently, each livestream is likely to have a substantial number of viewers who can serve as potential ``labelers'' to flag fanchuan users. Thus, we believe that for most of the fanchuan instances, we have ``labelers'' for them. Conversely, if there are fewer viewers, any fanchuan behavior that might occur would have a limited impact, thereby minimizing the consequences on our analysis. 
}

\pb{\blue{Do Labelers Identify all Fanchuan Behavior?}}
\blue{Fanchuan behavior can be flagged by multiple viewers, and as long as one viewer flags it, it will be successfully identified and included in our dataset. As a result, we indeed get a large number (130k) of fanchuan instances through this method. We acknowledge that there may be some perfect instances of fanchuan that manage to deceive all viewers and go unflagged. In such cases, it would be impossible to identify: if none of the (thousands of) viewers, who possess knowledge and experience of the context, can detect it, it is unlikely that external researchers would be able to identify it.}

\pb{\blue{Correct Keyword Selection?}}
\blue{The three selected keywords represent well-established conventions within the context of bilibili and fanchuan culture. Messages intended to flag fanchuan behavior are meant to be understood by all viewers, so they tend not to include expressions that could be misunderstood. 
The Chinese language also makes keyword search more robust, compared to English language text.
This is because variations of the same word (\eg with different stems) will still match the keywords, \ie the common Chinese characters remain the same even when appearing within modified or contextualized phrases. For example, in English, a naive match for the word ``troll'' may miss nuanced words like ``trolls'' and ``trolling'', but this is not the case in Chinese.}

\subsection{Data Preprocessing}
\label{subsec:nlp_methods}

\pb{Text Embedding.}
In some of our analyses, we utilize text embedding methods. To transform chats into vectors, we employ the \texttt{text2vec} project \cite{Text2vec} and opt for the \texttt{text2vec-base-chinese} model, known for its superior performance for Chinese language text~\cite{Text2vec, embeddingTest}.

\pb{Sentiment Analysis.}
In some analyses, we inspect the sentiment of chat messages. To accurately capture the distinct characteristics of Chinese Internet language, we utilize the SMP2020 dataset \cite{SMP2020}, which comprises Weibo posts annotated with six different sentiment categories (happy, angry, sad, fear, surprise, neutral). We use the \texttt{bert-base-chinese} model and fine-tune it on this dataset. 
\blue{Human verification is conducted to ensure the accuracy of the results. 
Specifically, we manually annotate 200 samples of chat messages to assess the fine-tuned model. The model achieved a macro F1-score of 73\% on the samples. This confirms the effectiveness of the model.}

\pb{Toxicity of Chats.}
We later explore the toxicity present in chat messages. To quantify this toxicity, we employ a fine-tuned COLD model \cite{Toxicity-Detection} introduced in \cite{Jiang_Shen_Wen_Sha_Chu_Liu_Backes_Zhang_2024}. This model is specifically fine-tuned for Bilibili bullet chats and is reported to achieve significantly better results compared to other mainstream tools. The classification result of this model is binary, and the chat is marked as 0 (non-toxic) and 1 (toxic).

\subsection{Ethical Considerations}
The data used in this study is openly available to any Bilibili user. We gather bullet chats and other messages publicly displayed during live streaming sessions on the Bilibili platform. This data is designed to be openly visible to any livestream viewer. We collect the sender's ID, the content of the message, and the specific timestamp.
We note that we do not and cannot link the user IDs on Bilibili with actual personal identities. Our analysis does not focus on individual users; instead, we aggregate the data to obtain a broader understanding.
For certain archived livestreaming sessions, we extract data from \texttt{danmakus.com}, a website that archives records of Bilibili livestreaming sessions. The website states that the data can be used for research purposes, with the condition that the source is credited.
We have obtained approval from the Institutional Review Board (IRB) at our home institution for this project.

\section{Characterization of Fanchuan (RQ1)}
\label{sec:RQ1}

\begin{figure}
     \centering
     \includegraphics[width=.5\linewidth]{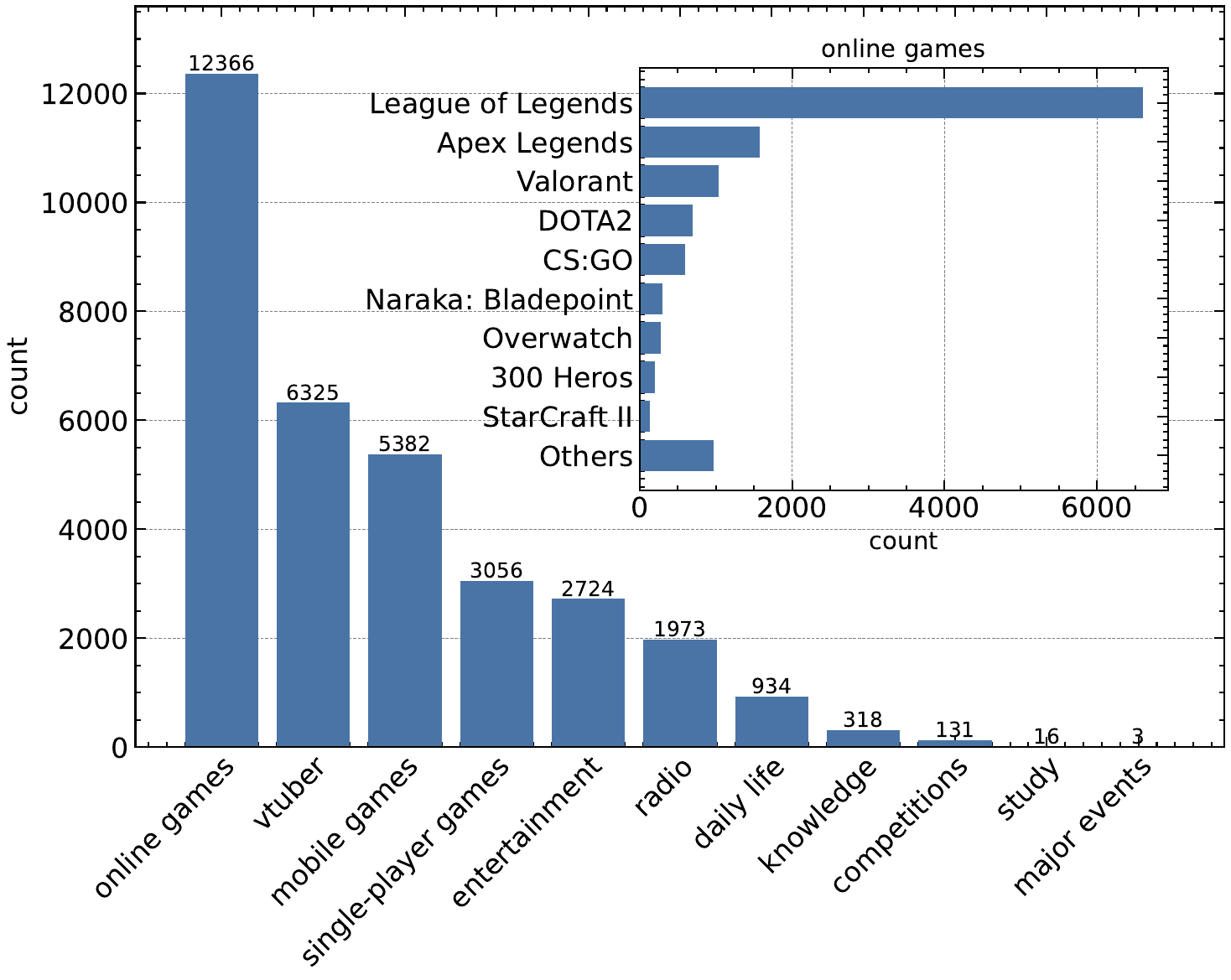}\hfill
     \includegraphics[width=.5\linewidth]{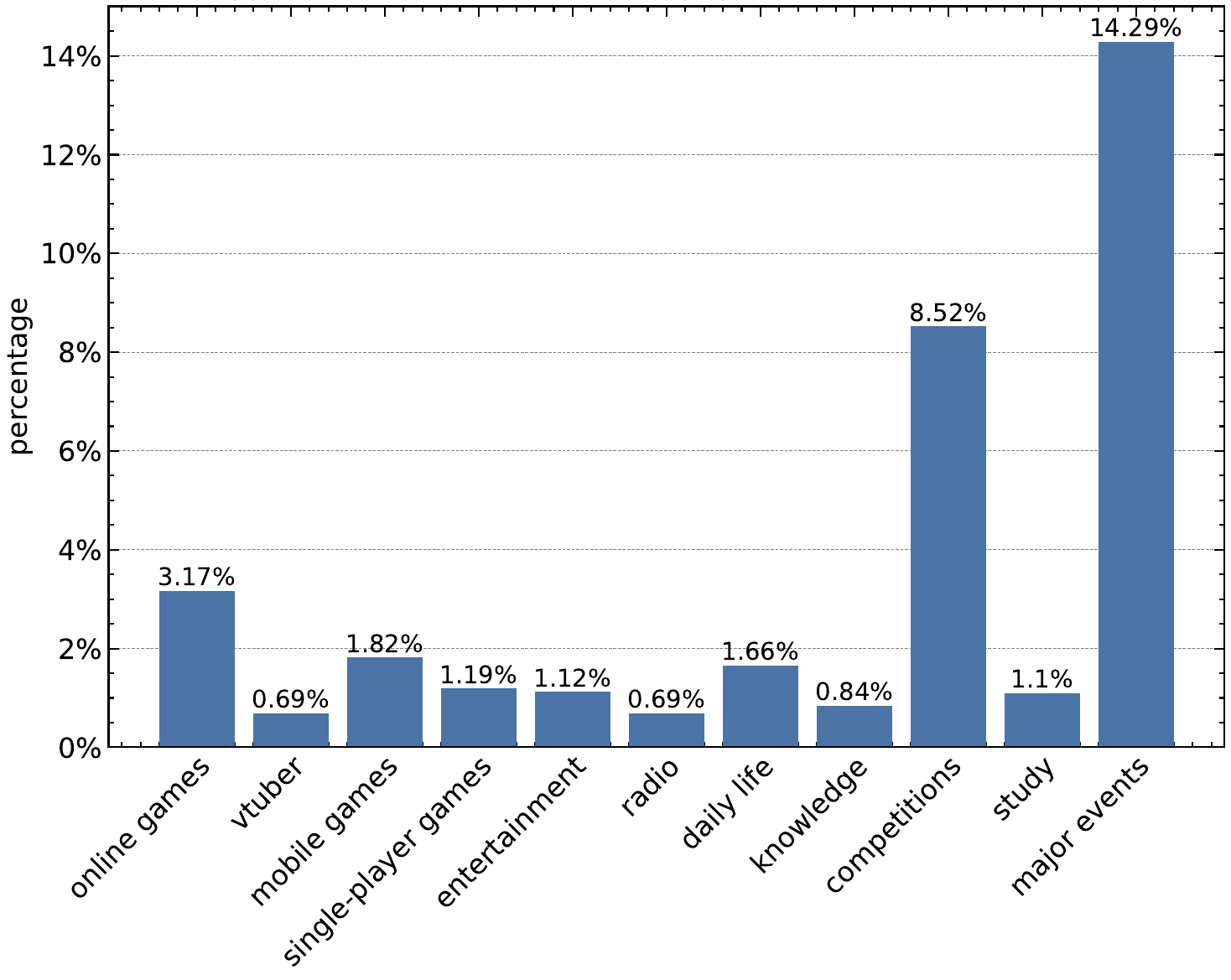}\hfill
     \caption{(a) Distribution of livestreaming area containing fanchuan behavior. (b) Percentage of livestreamings contain one or more fanchuan behaviors in each area.}
     \label{fig:rq1_1}
\end{figure}

\subsection{Characterization of Livestreaming with Fanchuan Chat}
\label{sec:RQ1_1}
\pb{\minorblue{Methodology.}}
We first examine the distribution of fanchuan chat messages within various livestreaming areas. We extract the parent area information of all livestreams, where a fanchuan chat exists. There are a total of 11 distinct areas.

\pb{\minorblue{Result.}}
Figure \ref{fig:rq1_1}a shows the prevalence of fanchuan chat across these different areas.
It becomes evident that gaming-related (online games, mobile games, single-player games) livestreams emerge as the predominant arena for fanchuan messages.
Specifically, online gaming livestreams represent the largest share, constituting 33.03\% of all livestreams that include fanchuan chat messages.
To dive deeper into this phenomenon, we further dissect the online gaming category, as shown in a sub-graph within Figure \ref{fig:rq1_1}a. The results reveal that \textit{League of Legends} outpaces all other games in terms of fanchuan chat volume.
One potential explanation is that \textit{League of Legends}, a globally recognized competitive online game, has a large and loyal player base. As with competitive sports, players usually support one or more professional players or teams. The intensely competitive nature of the game likely encourages people to exhibit antisocial behavior.
We further explore the entities targeted by fanchuan behavior in  \S\ref{sec:RQ1_2}.

While online games emerges as a prominent area for fanchuan behavior, it also boasts the largest volume of livestreams in our dataset (13.90\%). 
Thus, to normalize the results, Figure \ref{fig:rq1_1}b plots the histogram of the percentage of streams that contain one or more fanchuan message per area.
The results indicate that online games is indeed a \blue{clear} hotspot for fanchuan activity, \ie fanchuan behavior can be found in 3.17\% of the livestreams in the online games area, which is higher than all other areas with a relatively large number of livestreams. 
In contrast, while the VTuber area hosts a large number of livestreams targeted by fanchuan, the overall proportion is small (0.69\%).
The competition area is also a hotspot (8.52\%). We argue that certain specific entity associated with sports competitions may become the focus of fanchuan activity, thereby resulted in the frequent occurrence of fanchuan attacks.
Thus, in the next subsection, we further explore the target of the fanchuan attack.

\subsection{Targets of Fanchuan Chat Messages}
\label{sec:RQ1_2}
We next examine the target entities in the fanchuan chat messages. This analysis helps us understand the types of entities commonly targeted by fanchuan users, or the specific topics these fanchuan chat messages frequently focus on.

\pb{Methodology.}
We use the data in the Fanchuan User List described in \S\ref{subsec:dataset_construction}, as this analysis requires the exact fanchuan chat messages.
To extract entities from the chat messages, we utilize OpenAI's GPT-4o-mini model, employing a specific prompt detailed in Appendix \ref{subsec:appendix_prompts_nouns}. This gives all nouns mentioned in a chat message. In total, we extract 12,675 nouns from the fanchuan messages identified in \S\ref{subsec:dataset_construction}. Subsequently, we manually eliminate 7845 (61.9\%) nouns that do not refer to a specific entity (\eg ``I'', ``Computer'', ``Player''). After this filtering, we categorize the remaining nouns into different groups through manual labeling.

\pb{Result.}
The results are presented in Table \ref{tab:fanchuan_entity}.
We categorize these entities into five main areas: \blue{Professional} Esport, Streamer, Mobile/Gacha Game, \blue{Professional} Sport, and Miscellaneous. 
The highest proportion of fanchuan messages target Esport-related entities, making up 54.06\% of the total. This underscores the fact that professional gamers and teams are often the focus of fanchuan behavior, aligning with our findings in \S\ref{sec:RQ1_1} and previous studies that have documented the high levels of toxicity within esports livestreaming \cite{Jiang_Shen_Wen_Sha_Chu_Liu_Backes_Zhang_2024}.

Streamers, too, encounter a \blue{high} volume of fanchuan messages, comprising 21.07\% of the mentions. This suggests that the personal nature of streaming may render individuals particularly susceptible to attacks. While prior research has indicated that streamers frequently face online harassment, these instances are direct \cite{10.1145/3452918.3458794, 10.1145/3610191}. The phenomenon of using fanchuan to target one streamer from the audience of another streamer's session appears to be under-researched. This highlights the need for more sophisticated mechanisms to safeguard streamers' online safety.

Mobile and gacha games, such as Genshin Impact, account for another large portion (20.90\%) of the context for fanchuan behavior. Surprisingly, this has not been extensively discussed in previous research \cite{10.1145/3611043, 10.1145/3411764.3445157, 10.1145/3474680}. 
Intuitively, gacha games, being minimally competitive with no direct player-versus-player conflict or formal esport events, would be expected to foster a more peaceful community .
Traditional sports figures and teams, including those in football and basketball, also experience fanchuan messages, which represent 12.27\% of the total. This reveals that even outside of digital spaces, competitive discussions can devolve into hostility.
Lastly, a diverse category labeled miscellaneous accounts for 18.37\% of fanchuan messages. This covers a broad array of entities including nationalities, companies, and social behaviors. \blue{This suggests that fanchuan behavior may potentially stem from various factors} including gender, regional, national, and brand loyalties.

Overall, the findings show that the target of fanchuan attacks span a wide range of different areas and online communities. 
While it targets common areas where there tend to be a high conflict and toxicity like esports, it also extends its reach to other areas that might initially seem less prone to such behaviors. 
This diversity and adaptability of fanchuan behavior across various online communities underscore the complex challenge it poses.

\begin{table}[t]
\centering
\resizebox{\linewidth}{!}{%
    \begin{tabular}{|R{6em}|C{2.6em}L{26em}L{26em}|}
\toprule
    \textbf{Category} & \textbf{\%} & \textbf{Representatives} & \textbf{Description} \\
\midrule
    \textbf{\blue{Professional} Esport} & 54.06 & \zh{uzi/污渍/物资/乌兹 (player), ad/AD (in-game position), 大B/Doinb (player), EDG (club), 小孩 (player)} & \blue{Professional} esports players, clubs, in-game positions (usually also used to implicitly refer to some specific players). \\\midrule
    \textbf{Streamer} & 21.07 & \zh{孙亚/孙亚龙 (Esport streamer), DYS/德云色 (Esport streamer group), 梓神 (Chinese VTuber), 猫雷 (Japanese VTuber)} & Various kinds of streamers from different areas. \\\midrule
    \textbf{Mobile / Gacha Game} & 20.90 & \zh{原神 (Genshin Impact, game), 幻塔 (Tower of Fantasy, game), 米/米哈游/mhy (miHoYo, company), 鸣潮 (Wuthering Waves, game), OP (slurs towards Genshin player), 散兵 (character in Genshin)} & Mobile games and Gacha \cite{10.1145/3579438} games, and the related in-game entities or the company operating the game. \\\midrule
    \textbf{\blue{Professional} Sports} & 12.27 & \zh{梅西 (Messi, player),  巴萨 (Barcelona, club), 阿根廷 (Argentina), C罗 (Cristiano Ronaldo, player),  阿伟罗 (slurs towards Cristiano Ronaldo), 皇马 (Real Madrid, club),  猩猩 (slurs towards LeBron James), 詹姆斯 (LeBron James, player)} & \blue{Professional} sports players and clubs, such as in football or basketball. \\\midrule
    \textbf{Misc} & 18.37 & \zh{沸羊羊/舔狗 (simp), 中国人 (Chinese),  韩国人/韩国 (Korean/Korea), 河南 (province in China), 特斯拉 (Tesla), 华为 (Huawei)} & Common areas where hate speech and harassment tend to arise, including gender, region, and nationality. Additionally, industries where brand loyalty plays a significant role, such as cars, phones, and hardware. \\\midrule
\bottomrule
\end{tabular}%
}
\caption{Categories of entities presented in fanchuan chat messages.}
\label{tab:fanchuan_entity}
\end{table}

\section{Impact of Fanchuan on Livestreams (RQ2)}
\label{sec:RQ2}

In this section, we examine how fanchuan behaviour affects livestreams, focusing on user chat quantity, chat sentiment and toxicity. By analyzing these factors, we gain insight into the potential harm and impact of fanchuan behavior on livestreaming.

\subsection{Impact of Fanchuan on Chat Quantity}
\label{sec:rq2_1}

\begin{figure}
     \centering
     \includegraphics[width=.51\linewidth]{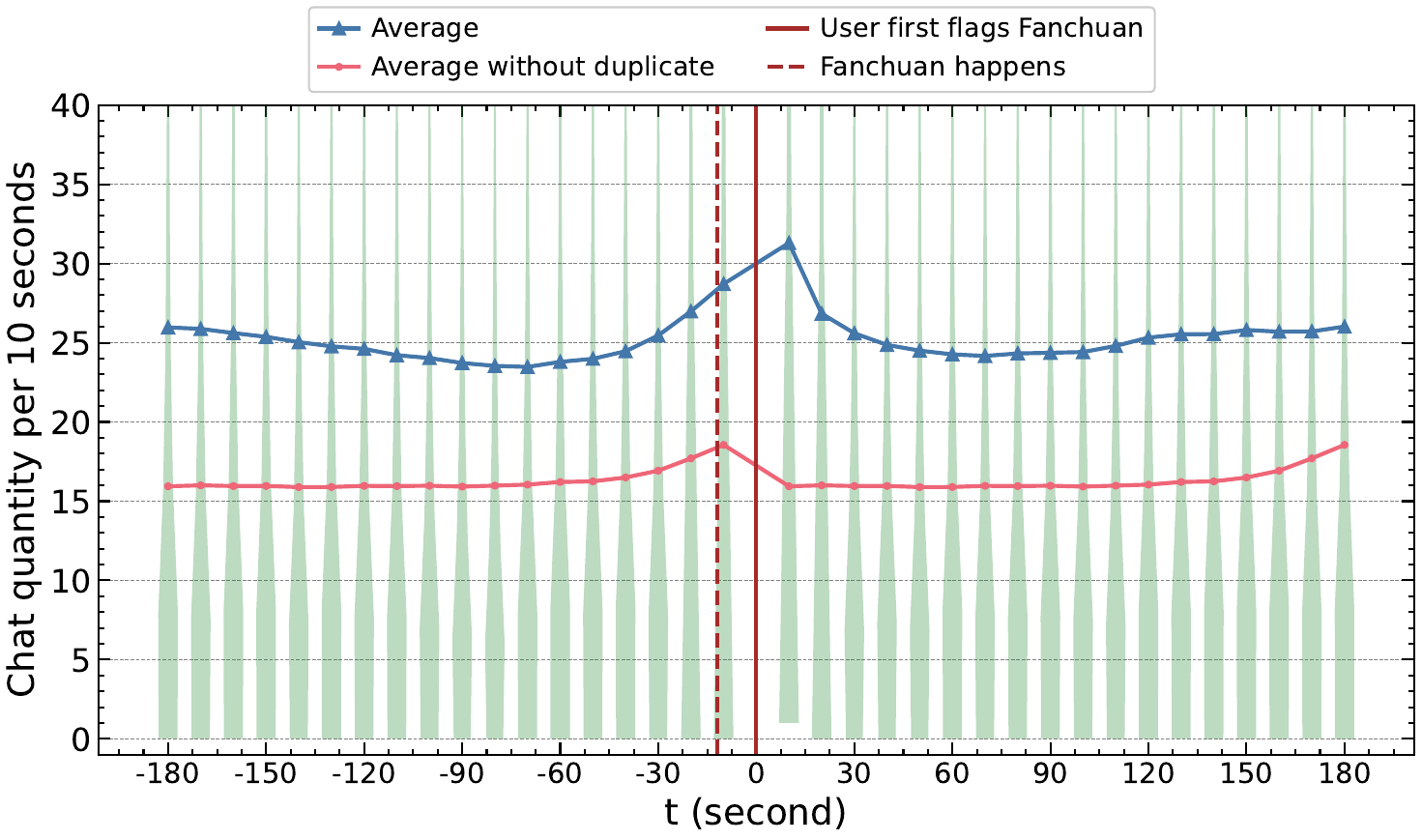}\hfill
     \includegraphics[width=.49\linewidth]{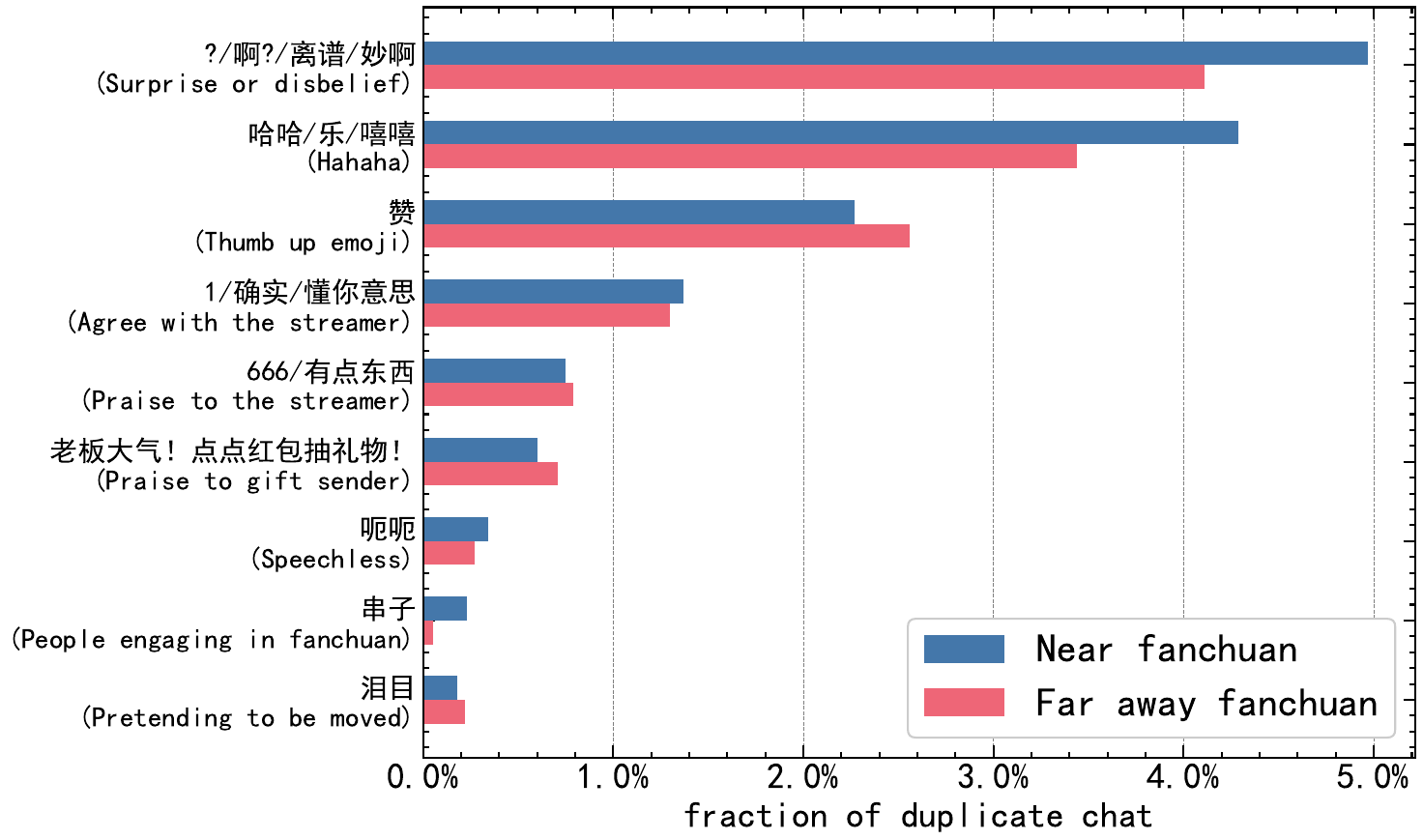}\hfill
     \caption{(a) Chat quantity in the three minutes before and after a user's first flag fanchuan. (b) Top 9 merged duplicate chat message from three minutes before and after a user's first flag fanchuan.}
     \label{fig:rq2_1}
\end{figure}

\pb{Methodology.}
We first examine the impact of fanchuan behavior on user chat quantity in livestreaming to quantify user engagement during the period where fanchuan happens.
For this purpose, our initial step involves pinpointing the time where fanchuan behavior takes place. 
For all 37,435 livestreamings containing (at least one) fanchuan behavior, we extract the timestamp when users first flag the fanchuan behavior. 
Then, using the timestamp as a reference point, we extract the users' chat quantity in 10 seconds intervals, covering three minutes before and after the timestamp. In cases where multiple fanchuan activities are detected within a single livestream, we use 30-seconds as a threshold to separate different fanchuan behaviors.
\blue{As a supplement, we test different time intervals, including 60-seconds, 90-seconds and 120-seconds. Through manual inspection, we find that 30-seconds intervals perform best in distinguishing different fanchuan behaviors, and longer time intervals may lead to misidentifying different fanchuan behaviors as the same instance.}

For reference, we also estimate the exact time when the fanchuan chat message is sent.
Recall, we can only identify the accurate timestamp of a fanchuan chat message for data in the Fanchuan User List as described in \S\ref{subsec:dataset_construction}.
\blue{To do this, we calculate the timestamp difference between the fanchuan chat message and the first chat message that flags fanchuan for all cases in the Fanchuan User List. As a result, we find the average timestamp difference is 11.96 seconds. Thus, we use T-11.96 as the estimated time point for the fanchuan behavior.}

\pb{Result.}
Figure \ref{fig:rq2_1}a shows the \blue{average} user chat quantity in the three minutes before and after the users' first flag fanchuan ($t=0$). \blue{Please note that each data point represents the average value over a 10-second interval, rather than the average at a specific time point.}
We observe a significant increase in the quantity of chat messages near to the user's first flag of fanchuan ($t=0$). However, this impact does not last for a long period (only about 2 minutes).
This suggests that there is a high level and short-term of user engagement with the fanchuan behavior, and that the fanchuan behavior may be a key factor that motivate users to chat.
We suspect that the surge in user chat quantity in a short period of time is due to duplicate chat, which is a common phenomenon on Bilibili \blue{(\ie users tend to imitate interesting chat messages sent by other users to increase their chances of being noticed by the streamer)}.
For further verification, we drop duplicate chat message for each 10 second intervals, and the unique chat quantity results are shown in the red line in Figure \ref{fig:rq2_1}a. 
Surprisingly, even though the chat quantity surges at $t=0$ (timestamp of user first flag fanchuan), the unique chat quantity decreases after removing duplicate chats.
This suggests that there are many duplicate chats and that users might follow trends in sending chats (\ie repeating the same short message).

To further analyse whether the duplicate chats are related to fanchuan, we extract all the chat messages during the two time periods:
\begin{enumerate}
    \item \textbf{Close to fanchuan period}, $-10 \leq t \leq 10$, around the time when chat quantity surges.
    \item \textbf{Far from fanchuan period}, $-180 \leq t \leq - 170 \text{ and } 170 \leq t \leq 180$.
\end{enumerate}
\blue{Recall, we find that the time between fanchuan behavior and the first chat mention fanchuan differs by 11.96 seconds. Therefore, the \texttt{close to fanchuan period} should include $-10 \leq t$, since this window likely captures potential fanchuan-related activity.}
Next, we extract the top 30 most frequent duplicate chats from \texttt{close to fanchuan period}, and merge similar chats through manual checks. We then calculate the fraction of each merged duplicate chat in the \texttt{far from fanchuan period}. We argue that comparing the differences in chat messages between the two periods helps us understand which chats are driven by fanchuan behavior.

Figure \ref{fig:rq2_1}b shows the top 9 merged duplicate chat messages from \texttt{close to fanchuan period} (blue) and \texttt{far from fanchuan period} (red).
The most common duplicate chat messages are \zh{?/啊?/离谱/妙啊} (Surprise or disbelief) and \zh{哈哈/乐/嘻嘻} (Hahaha, may also be used for mockery), accounting for 4.97\% and 4.29\% of all chats for \texttt{close to fanchuan period}, and are 20.92\% and 24.71\% higher than same chats in the \texttt{far from fanchuan period}.
The fractions of these two terms differ between the two time periods, suggesting a change in user sentiment around the fanchuan behavior. 
In contrast, we also observe that the fraction of \zh{串子} (People engaging in fanchuan) during the \texttt{close to fanchuan period} is 360\% higher than during the \texttt{far from fanchuan period}. 
\blue{This suggests that near fanchuan behavior, users are more inclined to discuss those who are engaging in the fanchuan.}
\blue{As a supplement, we extract the top 60 most frequent duplicate chats (from \texttt{close to fanchuan period}) and redo the same experiment above. The results show no additional fanchuan-related terms. Moreover, since many of these duplicate chats account for only a very small proportion, we cannot confidently attribute them to fanchuan behavior.}

Overall, the findings suggest that there is a significant increase in chat quantity around the fanchuan behavior, and that fanchuan behavior is one of the factors contributing to the surge in chat quantity. 
As fanchuan can be considered a toxic behavior, we conjecture that this surge may also be driven by the increase of chat messages with negative sentiment or toxicity. Thus we further explore this in the next subsection (\S\ref{fig:rq2_2}).

\subsection{Impact of Fanchuan on Chat Sentiment \& Toxicity}
\label{sec:rq2_2}

\begin{figure}
     \centering
     \includegraphics[width=.485\linewidth]{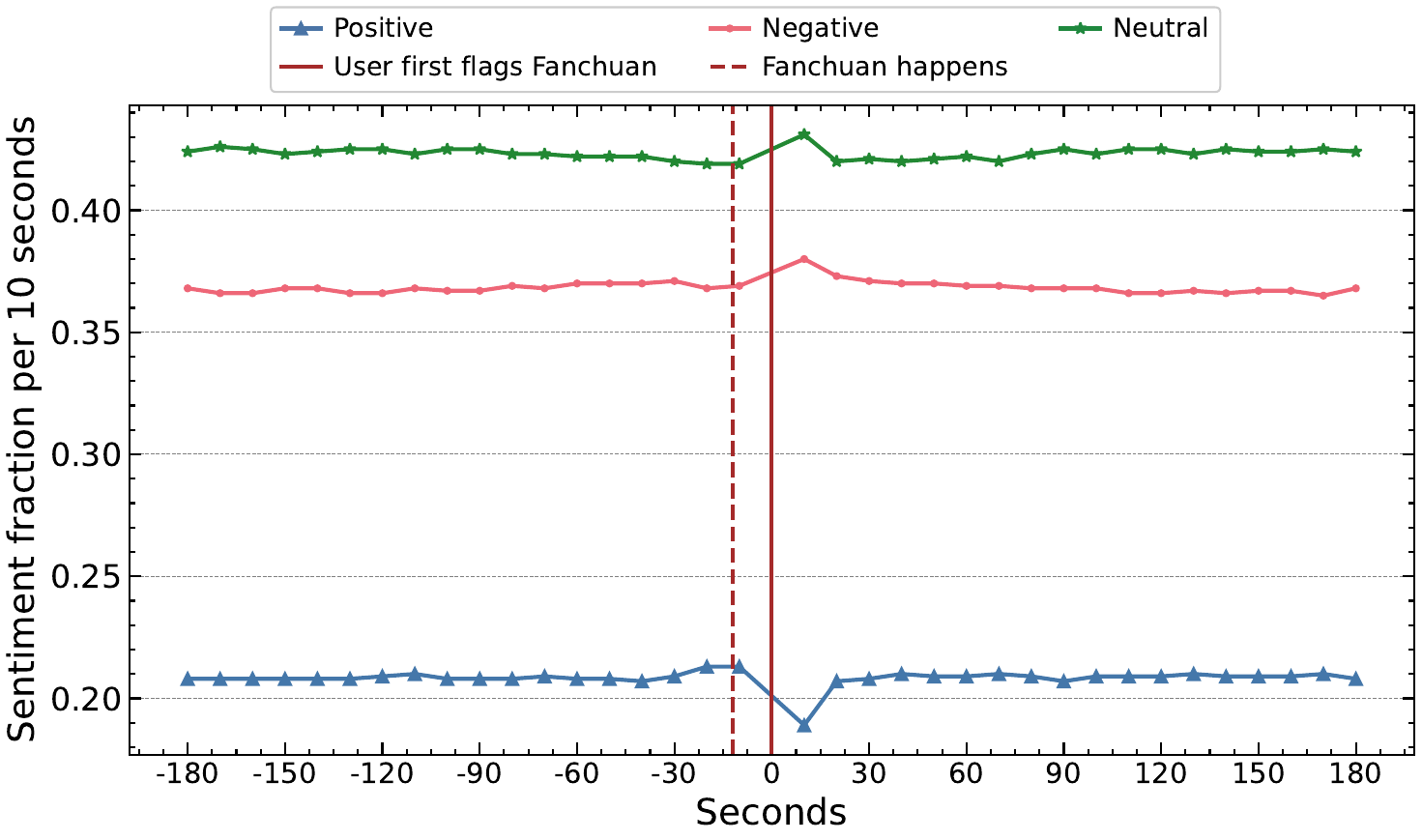}\hfill
     \includegraphics[width=.505\linewidth]{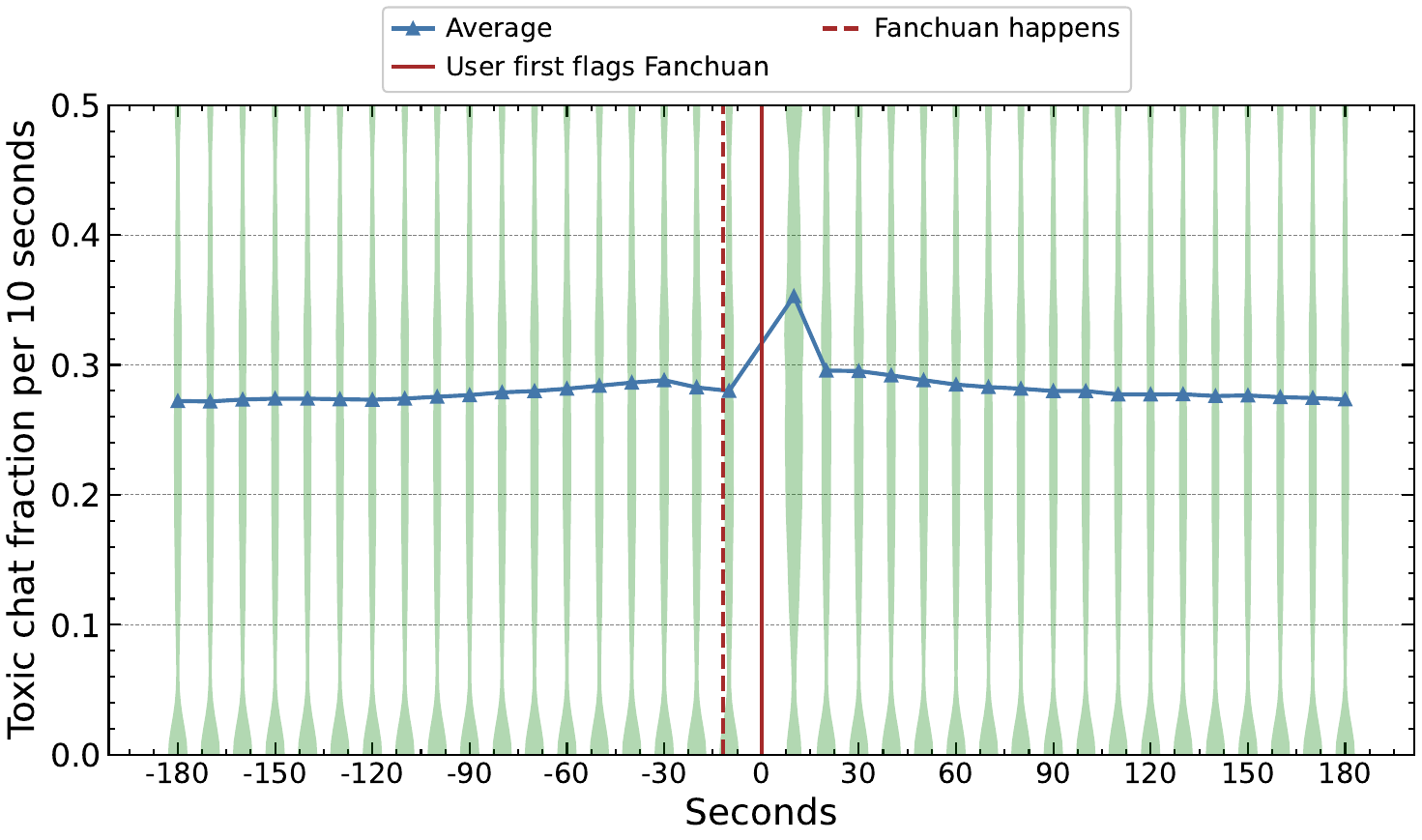}\hfill
     \caption{(a) Users' chat sentiment in the three minutes before and after a user's first flag fanchuan. (b) Users' chat toxicity in the three minutes before and after a user's first flag fanchuan.}
     \label{fig:rq2_2}
\end{figure}

To quantify the harmful effects of fanchuan, we further measure the changes in users' chat sentiment and toxicity around fanchuan behaviors.
We argue that this helps us better understand how the harmful effects of fanchuan spread in user chats, and how long they persist for, providing insights for fanchuan moderation strategies.

\pb{Methodology.}
As in \S\ref{sec:rq2_1}, we use the timestamp of the user's first fanchuan as a reference point ($t=0$), and then use the pre-trained model to classify the sentiment and toxicity of all chats three minutes before and after that timestamp. Please refer to \S\ref{subsec:nlp_methods} for the technical details.

\pb{Result.}
Figure \ref{fig:rq2_2}a shows how the three types of sentiment (positive, negative, and neutral) are distributed in chats concerning the fanchuan behavior. 
Overall, neutral and negative chats account for a larger fraction, exceeding positive chats by 27.4\% and 15.5\%, respectively.
In addition, there is an increase in negative and neutral chats near the user's first flag fanchuan (close to fanchuan period, $-10 \leq t \leq 10$, see \S\ref{sec:rq2_1}), while the fraction of positive chats decreases.
This suggests that there is a clear change from positive to negative and neutral sentiment in user chat around fanchuan behavior.
The increase for neutral chat is intuitive. Recall that there is an increase in chat quantity around the fanchuan (see \S\ref{sec:rq2_1}) and there is a large amount of duplicate chat (see Figure \ref{fig:rq2_1}b).
Among them, many duplicate chats are neutral sentiment (33\% of the top 100 repeated chats are identified as neutral chats, \eg \zh{属实离谱 (That is ridiculous), \zh{懂你意思 (Agree with you)}}), which leads to an increase in the fraction of neutral chats near the fanchuan behavior.

At the same time, the number of negative sentiment chats are also increased,
which may be influenced by the harmful effects of fanchuan behavior.
To further verify, we check the chat toxicity during the same time period (three minutes before and after the user's first flag fanchuan), aiming to quantify the toxicity of negative chat. The time series results are shown in Figure \ref{fig:rq2_2}b.
Not surprisingly, the fraction of users' toxic chat has the same growth trend as negative chat, and the fraction of toxic chat peaks at $t=10$. 
Unlike the chat quantity, the impact of fanchuan on the user's chat sentiment and toxicity lasts less time, about 30 seconds (from $t=-10$ to $t=20$).

Overall, fanchuan behavior may have an immediate impact on the sentiment of users' chats. Near the fanchuan behavior, the negativity and toxicity of chat messages increases significantly \blue{(with p-value \textless 0.001 when comparing the average quantity of negative or toxic chats \texttt{close to fanchuan} \vs \texttt{far from fanchuan})}.
Thus, we further investigate whether the rise in toxic chats is directly caused by fanchuan behaviors.

\subsection{Topic Analysis.}

\pb{Methodology.}
To further examine whether toxic chat content is directly related to fanchuan, we extract all toxic chats during the \texttt{close to fanchuan period} (which is the time period when toxic chats surge) and \texttt{far from fanchuan period}, respectively. We then use BERTopic \cite{grootendorst2022bertopic} to train two separate topic models on the extracted chat in each period (1,018,866 chats in \texttt{close to fanchuan period}, 921,353 chats in \texttt{far from fanchuan period}).
BERTopic employs sentence-transformers \cite{reimers2019sentence}
and c-TF-IDF \cite{grootendorst2022bertopic} to generate compact clusters of information, facilitating the interpretation of topics while retaining significant words within the topic descriptions.

\pb{Results.}
We identify a total of 172 topics in the \texttt{close to fanchuan period} and 152 topics in the \texttt{far from fanchuan period}. 
Table \ref{tab:toxicity_topic} shows the distribution of the top 10 topics in the two periods. We manually label each topic with its general umbrella category.
We observe that some topics (excluding outlier) have considerable differences between the two periods. The most obvious is topic 0 (topic related to fanchuan behavior and fanchuan users), which is 3.48\% in \texttt{close to fanchuan period}, which is 640.4\% higher compared to \texttt{far from fanchuan period}.
This suggests that fanchuan behaviors and fanchuan users are directly responsible for the significant increase in toxic chats.
Further, topic 4 (an esport streamer), topic 5 (pointing out the fanchuan behavior on bullet chat) and topic 7 (trolling behavior) are 78.2\%, 43.1\% and 37.7\% higher in \texttt{close to fanchuan period} compared to \texttt{far from fanchuan period} respectively.
These topics include common target entities to fanchuan behavior (see \S\ref{sec:RQ1_2} and Table \ref{tab:fanchuan_entity}). Although they are also discussed during \texttt{far from fanchuan period}, the significant increase in engagement in \texttt{close to fanchuan period} suggests that users' toxic chat are not limited to fanchuan behavior or users, but extend to target entities of the fanchuan behavior.

Overall, fanchuan behavior is a key factor in the temporary surge of toxic chats. This includes toxic chats targeted at fanchuan behavior or users, and this ``attack'' also extends to other entities with the fanchuan behavior.
This highlight the harmfulness of fanchuan behavior and the necessity of moderation for fanchuan behavior.

\begin{table}[t]
\centering
\resizebox{\linewidth}{!}{%
    \begin{tabular}{|R{2em}|C{6em}C{6em}|L{26em}L{26em}|}
\toprule
    \textbf{Topic} & \textbf{Close to fanchuan period (\%)} & \textbf{Far from fanchuan period (\%)} & \textbf{Representatives} & \textbf{Description} \\\midrule
    
    \textbf{-1} & 50.92 & 54.10 & \zh{狙神 (terms in the game League of Legends), 笑 (laugh), !, 主播 （streamer）, 什么 (What)} & Outliers \\\midrule
    
    \textbf{0} & 3.48 & 0.47 & \zh{串子 (people engaging in fanchuan), 别串 (do not fanchuan), 串 (fanchuan)} & Some keywords about fanchuan behavior or fanchuan users \\\midrule
    
    \textbf{1} & 2.33 & 2.4 & \zh{主播 (streamer), 开播 (start a livestreaming), 直播间 (livestreaming room), 下播 (stop a livestreaming)} & Some malicious jokes that target streamer and current livestreamings \\\midrule
      
    \textbf{2} & 1.47 & 1.58 & \zh{赢 (win), 输 (lose), 打野 (in-game position), 打团 (group fight), 比赛 (competition)} & Some terminologies about esports competitions \\\midrule
    
    \textbf{3} & 0.98 & 0.84 & \zh{睡 (sleep), 别睡了 (do not sleep), 醒醒 (wake up)} & Usually used to mock streamers for being silent, making the stream too quiet, or to moke esports players for under-performing. \\\midrule
    
    \textbf{4} & 0.98 & 0.55 & \zh{孙亚 (esport streamer)} & An esport streamer \\\midrule
    
    \textbf{5} & 0.93 & 0.65 & \zh{弹幕 (bullet chat), 串 (fanchuan), 别串 (do not fanchuan)} & Point out the fanchuan behavior of certain bullet chats \\\midrule
    
    \textbf{6} & 0.90 & 0.92 & \zh{狗子 (dog), 当狗 (be a dog)} & Using a dog metaphor to insult an entity \\\midrule
    
    \textbf{7} & 0.84 & 0.61 & \zh{钓鱼 (fish), 别钓 (do not fish), 鱼塘 (fish pond)} & Some keywords about trolling behavior \\\midrule
    
    \textbf{8} & 0.79 & 0.73 & \zh{红包 (red envelope), 老板 (boss), 送礼物 (send gifts)} & Used in chat for lottery, either sent by the streamer or users. \\\midrule
    
    \textbf{9} & 0.75 & 0.75 & \zh{uzi (esport player), 韦鲁斯 (character in League of Legends)} & A famous League of Legends esport player \\\midrule
\bottomrule
\end{tabular}%
}
\caption{Distribution of the top 10 topics in \texttt{close to fanchuan period} and \texttt{far from fanchuan period}, and representative words for the topics.}
\label{tab:toxicity_topic}
\end{table}

\section{Characterizing Fanchuan Users (RQ3)}
\label{sec:RQ3}

In this section, we explore the characteristics of users who engage in fanchuan behavior. We contend that gaining insight into these characteristics can assist moderators and the design of moderation mechanism in identifying and preventing potential fanchuan instances. We note that for this RQ, our analysis focuses on the Fanchuan User List of 17,337 users in the constructed dataset as discussed in \S\ref{subsec:dataset_construction}.

\begin{figure}
    \centering
    \includegraphics[width=0.23\linewidth]{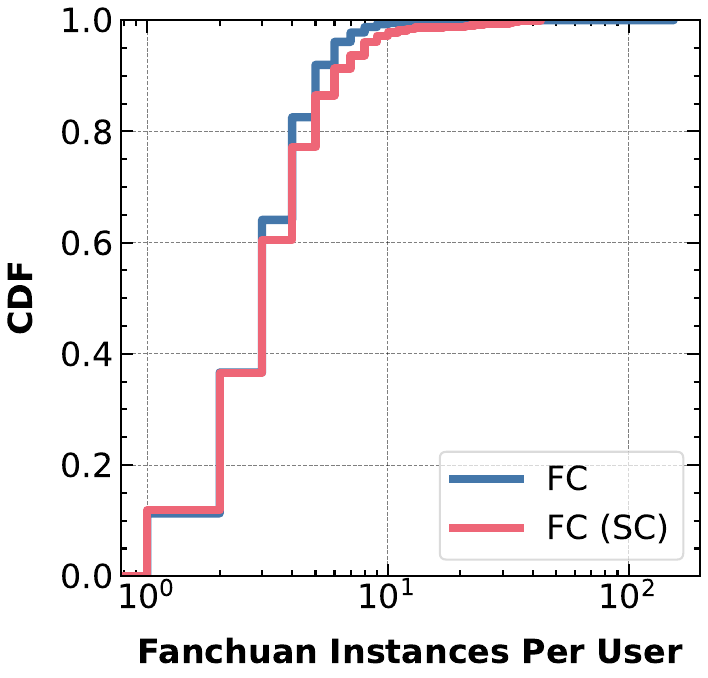}\hfill
    \includegraphics[width=0.23\linewidth]{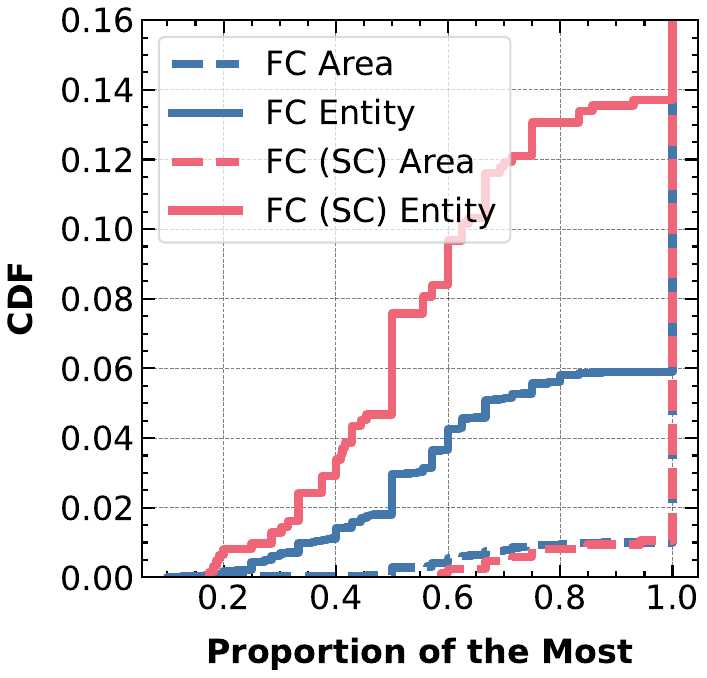}\hfill
    \includegraphics[width=0.23\linewidth]{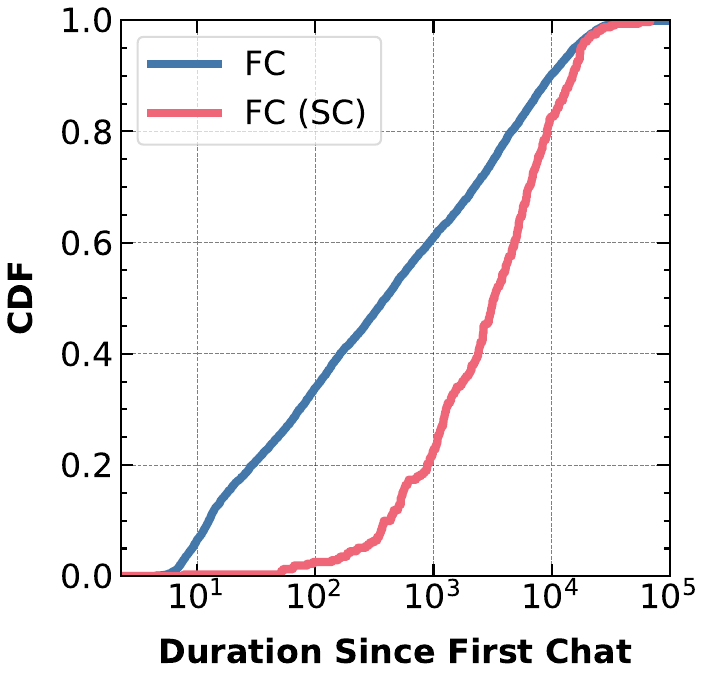}\hfill
    \includegraphics[width=0.23\linewidth]{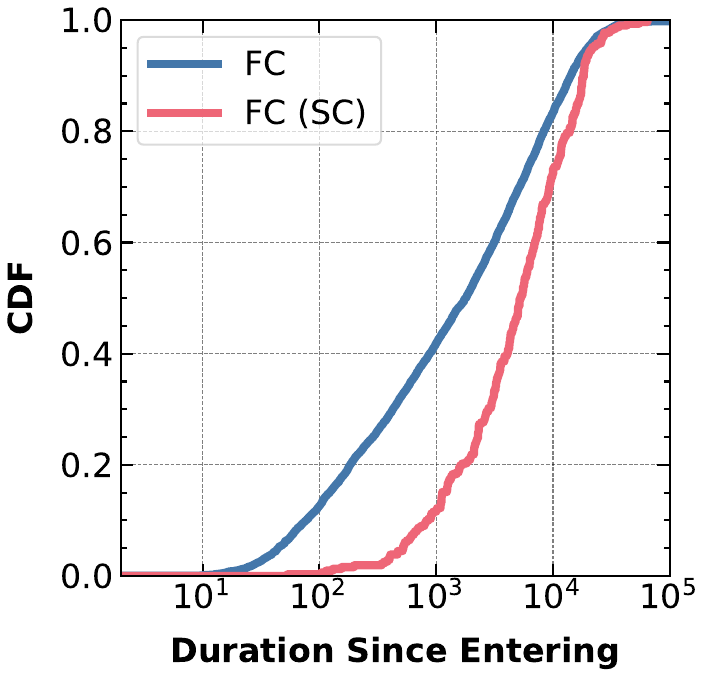}
    \caption{CDF of (a) the number of fanchuan instances per Fanchuan User; (b) the proportion of the most frequent area and entity for the Fanchuan User's fanchuan behavior; (c) the time (seconds) elapsed from the Fanchuan User's first chat message to their fanchuan chat message; (d) the time (seconds) elapsed from the Fanchuan User's entry to their fanchuan chat message.}
    \label{fig:rq3_1}
\end{figure}

\subsection{Quantifying the Patterns of Users' Fanchuan Behavior}
\label{subsec:rq3_1}

\pb{\minorblue{Methodology.}}
We first investigate the patterns of Fanchuan Users' fanchuan behavior. We posit that fanchuan activity is not merely a one-off random occurrence; rather, users engage in fanchuan on a regular basis with some patterns, which can assist in identifying potential fanchuan users.

\pb{\minorblue{Result: Number of Fanchuan Events Per User.}}
To start, we measure how often users engage in fanchuan actions. Figure \ref{fig:rq3_1}a presents the CDF of the number of fanchuan actions conducted by each fanchuan user. 
We see that most users are repeat offenders, with 88\% of the Fanchuan Users engaging in the activity more than once. Notably, the most active Fanchuan User participates in fanchuan over 100 times.

\pb{\minorblue{Result: Fanchuan Area \& Entity.}}
In \S\ref{sec:RQ1}, we find that fanchuan behaviors are typically directed towards a specific entity, and these entities vary \blue{greatly} across different genres of stream. Consequently, we hypothesize that each Fanchuan User concentrates their fanchuan activities on one particular area or entity.
To explore this hypothesis, we analyze the most frequently targeted areas and entities for each Fanchuan User. To quantify the results, we calculate the proportion of their most frequently targeted area and entity relative to all the areas and entities they target.
The resulting CDF is shown in Figure \ref{fig:rq3_1}b.
The findings confirm our hypothesis, revealing that the majority of fanchuan users indeed concentrate their efforts on a specific target. Specifically, 99\% of Fanchuan Users engage in activities within a single area, and 93\% focus on a single entity.

\pb{\minorblue{Result: The Time Point of Fanchuan.}}
Next, we turn our attention to identifying the specific time point within a livestreaming session when users exhibit fanchuan behavior. Specifically, we aim to determine the amount of time it takes for a user to perform fanchuan behavior after they begin watching the livestream. We wonder whether, similar to the patterns of random spammers \cite{IJAI20659}, a large proportion of fanchuan behaviors will occur shortly after the user begins watching.

Figures \ref{fig:rq3_1}c and \ref{fig:rq3_1}d display the CDF of the time (seconds) elapsed from the user's first chat message to their fanchuan behavior, and from the entry into the livestream to their fanchuan behavior. 
We see that only 30\% of fanchuan behaviors occur within the first 300 seconds after the Fanchuan User enters the livestream. Contrary to our initial expectations, this finding suggests that the majority of fanchuan activities take place after the users have spent a considerable amount of time watching the livestream. This is particularly notable in the case of fanchuan with superchats, where 99\% of the instances occur after the first 300 seconds.

To investigate this, we conduct a manual review of 200 sample messages (see Appendix \ref{subsec:appendix_manual} for details). This review confirms that fanchuan behaviors occurring shortly after the user starts watching are similar to random spammers, with the content being formulaic and only have some relation with the livestream's title or area.
In contrast, fanchuan messages sent after a longer viewing period tend to be more related to the livestreaming content or other chat messages. We conjecture that for these users, their fanchuan behavior may be triggered by something mentioned either by the streamer or in another viewer's chat.

Overall, the findings suggest that the majority of fanchuan users may, for most of the time, behave like regular users, indicating that traditional moderation mechanisms designed to combat bots and spammers might not be effective for fanchuan cases. This highlights the necessity for enhanced moderation strategies.

\begin{figure}
    \centering
    \includegraphics[width=\linewidth]{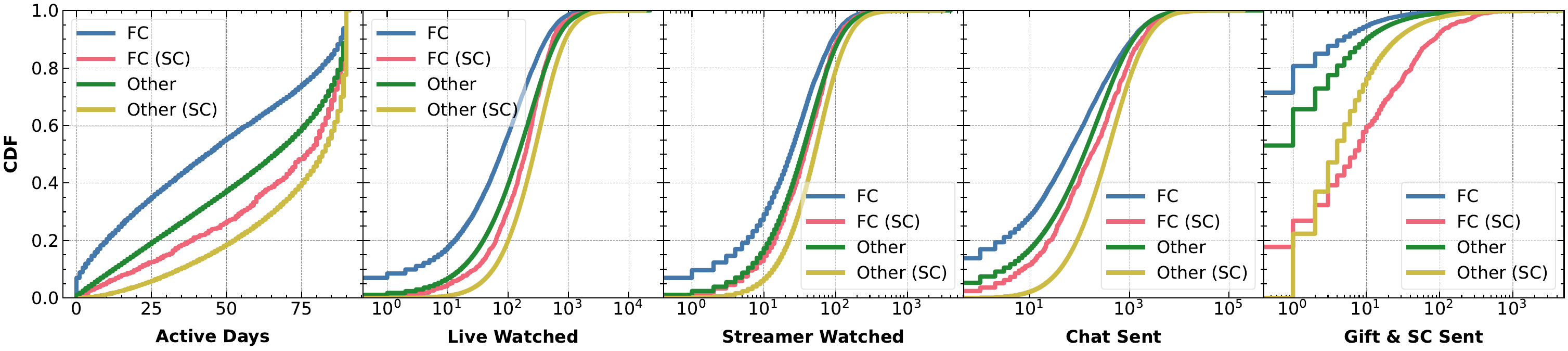}
    \caption{CDF of the activity metrics for Fanchuan Users and Comparison Users during the measurement period.}
    \label{fig:rq3_2}
\end{figure}

\subsection{Quantifying the Patterns of Users' Historical Activity}
\label{subsec:rq3_2}

Next, in order to further understand the fanchuan users, we inspect their platform-wide activities. We hypothesize that their historical activity (such as viewing, bullet chatting, and gifting) before their fanchuan behavior may differ from those of other users. 
Fanchuan can be generally considered as a toxic behavior. Thus, we posit that their messages may exhibit more negative sentiment and toxicity, even for livestreaming sessions where they may not send a fanchuan message.
These differences could aid in flagging potential fanchuan users.

\pb{Methodology.}
We use several activity metrics that have been widely used in previous studies of streaming platforms \cite{10.1145/3311350.3347149}.
Specifically, we employ 
    \one \textbf{Active days:} Number of active days of the user;
    \two \textbf{Livestream sessions watched:} Number of livestream sessions watched by the user;
    \three \textbf{Streamers watched:} Number of streamers watched by the user;
    \four \textbf{Number of chats:} Total number of chats sent by the user;
    and \five \textbf{Number of gifts and superchats:} Total number of gifts and superchats sent by the user.
    
For each Fanchuan User and Comparison User (from the Fanchuan User List and the Comparison User List in our dataset as described in \S\ref{subsec:dataset_construction}),
we calculate these metrics over a 90-day period. Specifically, for each user and the respective livestreaming session they perform fanchuan behavior, we consider a timeframe extending 90 days prior to the livestream session. This approach enables us to capture their behaviors before engaging in fanchuan, which can help in their identification beforehand.
We also measure the toxicity and sentiment of chat messages sent by users in the 90-day period before the fanchuan behavior. To evaluate the toxicity and sentiment, we employ the methods described in \S\ref{subsec:nlp_methods}.

As mentioned in \S\ref{subsec:dataset_construction}, a user may be counted multiple times if they are related to multiple instances of fanchuan, resulting in the metrics being calculated for multiple 90-day periods. For such users, we take the mean average of the metrics across these different 90-day periods.
We categorize Fanchuan Users into two groups: those who have used SC for their fanchuan activities, and those who have not. Likewise, for comparative purposes, we also categorize Comparison Users who have sent a SC during the 90-day period into a separate group.

\pb{\minorblue{Result: Viewing \& Chatting.}}
We first investigate user activity by looking at the number of active days, livestreaming sessions watched, unique streamers watched, and chat messages sent over the 90-day period. The results are displayed in Figures \ref{fig:rq3_2} (a-d).

Interestingly, we find that Fanchuan Users show less activity compared to Comparison Users. Specifically, Fanchuan Users are less active in all four metrics compared to Comparison Users within the same category. 
On average, a Fanchuan User is active for 45 days, watches 164 livestreaming sessions from 38 streamers, and sends 440 chat messages. 
In comparison, an average Comparison User is active for 58 days, watches 271 livestreaming sessions from 51 streamers, and sends 497 chat messages.

To understand this, we analyze the user profiles on Bilibili. We discover that 12\% of the Fanchuan User accounts have since been deleted. Account deletion on Bilibili is extremely uncommon, typically occurring only when the platform officially removes an account due to it being identified as malicious (\eg bots, spammers). For instance, fewer than 0.01\% of the corresponding Comparison User 
for these deleted Fanchuan Users are deleted.
Further investigation confirms that these users are the ones sending fanchuan messages shortly after (\ie less than 300s) entering the livestream, as discussed in \S\ref{subsec:rq3_1} (Figure \ref{fig:rq3_1}d). These individuals represent the least active segment of Fanchuan Users, participating for only one or even zero days (as shown in Figure \ref{fig:rq3_2}a) before likely being deleted.

However, this alone does not completely account for the lower levels of activity, particularly for Fanchuan (SC) Users.
As outlined in \S\ref{subsec:rq3_1}, Fanchuan Users tend to concentrate their fanchuan behavior in a specific area. Thus, we conjecture that, besides their fanchuan behavior, their overall activity is also focused on a specific area. That is, they predominantly watch livestreams and streamers within one or a few areas, leading to lower activity levels platform-wide.
To test this, we recalculate the metrics, focusing solely on the activity in the area where the Fanchuan User engaged in fanchuan behavior. 
The result as presented in Figure \ref{fig:rq3_activity_area} in the Appendix confirms the hypothesis, indicating that Fanchuan Users' activity within the specific area is comparable to that of Comparison Users.

Overall, the findings reveal distinct patterns of Fanchuan Users' activity, including viewing, chatting, and gifting, that \blue{greatly} deviate from those observed in comparison users. These differences could be instrumental in detecting potential fanchuan users for moderators and automatic moderation tools.

\pb{\minorblue{Result: Gift \& SC.}}
The activities of gift and SC differ from the above user activities because they involves money. Figure \ref{fig:rq3_2}e presents the CDF of the number of gifts and SCs sent by users over the 90-day period. We observe that Fanchuan Users \vs Comparison Users exhibit similar patterns to the above viewing and chatting activities. 
However, the trend flips for users sending SC, with Fanchuan (SC) Users sending more gifts and SCs compared to Comparison (SC) Users (average 35.4 \vs 16.3).
This tendency is intuitive, considering that Fanchuan (SC) Users, by spending a considerable amount of money to send a SC to fanchuan, demonstrate their strong willingness and financial capability to invest in the platform. 
These findings suggest that moderating Fanchuan (SC) Users presents a more intricate challenge. These users are not only active but also contribute substantially in financial terms, making them ``high-quality'' users whom the platform and streamers aim to satisfy.

\begin{figure}
    \centering
    \hfill
    \includegraphics[width=0.663\linewidth]{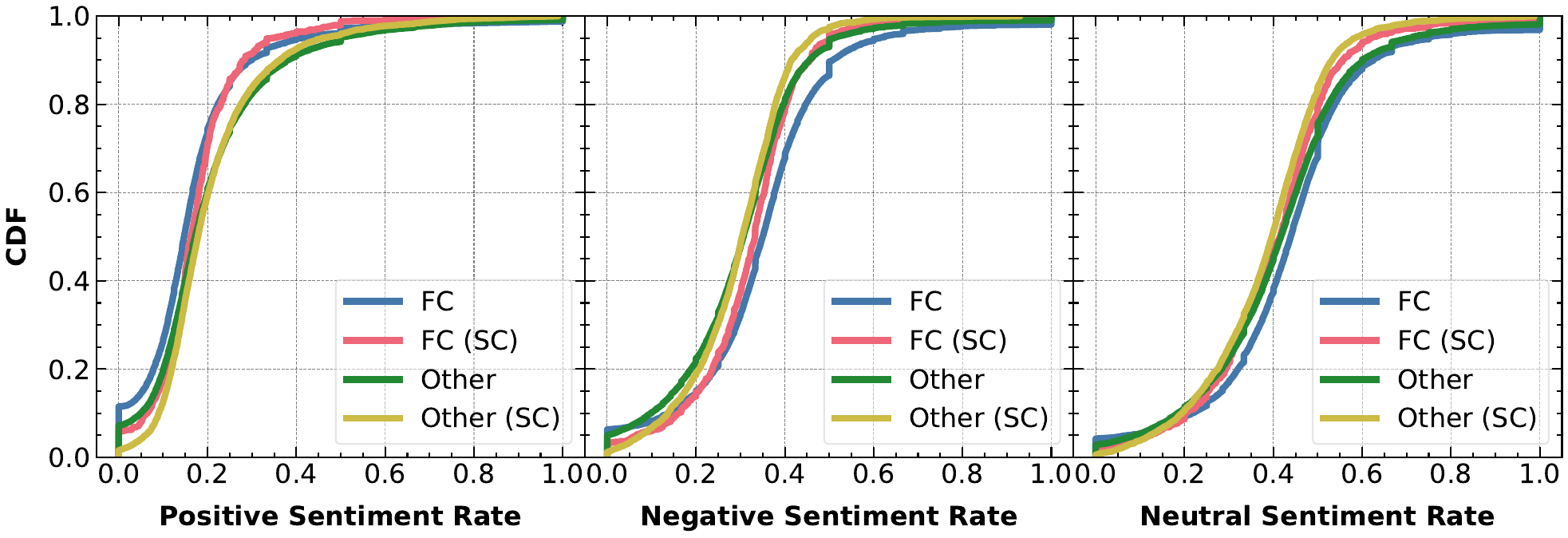}\hfill
    \includegraphics[width=0.255\linewidth]{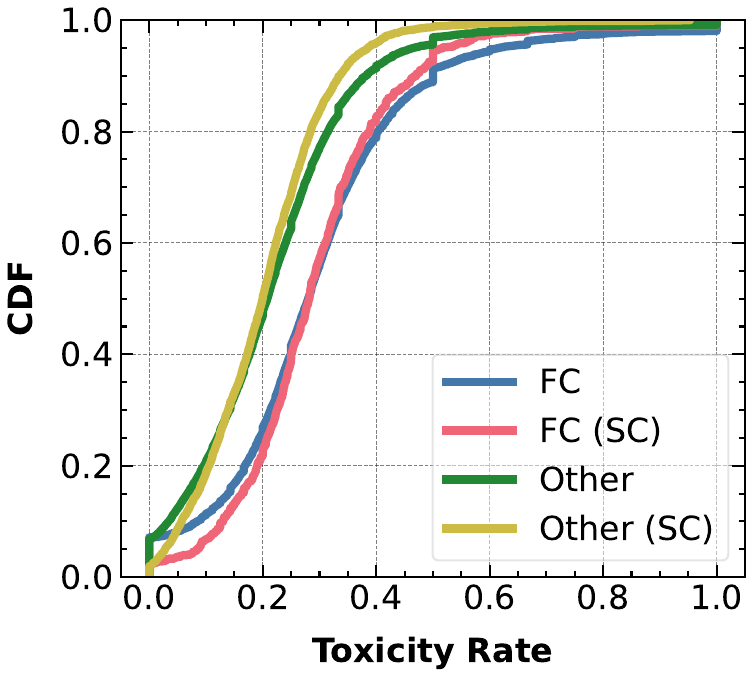}\hfill
    \caption{CDF of the proportion of chat messages with (a) positive, negative, and neutral sentiment; (b) toxicity, for Fanchuan Users and Comparison Users during the measurement period.}
    \label{fig:rq3_3}
\end{figure}

\pb{\minorblue{Result: Sentiment \& Toxicity.}}
Figure \ref{fig:rq3_3} (a-c) displays the CDF of the ratio of chat messages with positive, negative, and neutral sentiments, respectively, in relation to the total number of chat messages sent by a user.
Figure \ref{fig:rq3_3}d presents the CDF of the ratio of toxic chat messages in relation to the total chat messages sent by a user.

The findings corroborate our expectations, revealing that the share of chat messages with negative sentiments from Fanchuan Users surpasses that of Comparison Users, with the average of 35\% \vs 30\%. Conversely, the share of messages with positive sentiments is lower, with the average of 17\% \vs 21\%.
Delving deeper into the predominant emotion underlying the negative-sentiment chat messages from Fanchuan Users, we discover that the majority (79\%) are characterized by ``anger''. This indicates that Fanchuan Users are more prone to express their dissatisfaction through their chats, where engaging in problematic fanchuan behavior is also a way to show dissatisfaction.
As for the toxicity, the result is also in line with our expectations, demonstrating that the proportion of toxic chat messages from Fanchuan Users is indeed higher compared to Comparison Users, with the averages being 30\% \vs 22\%.
This suggests that Fanchuan Users are not only repeatedly involved in fanchuan, as indicated in \S\ref{subsec:rq3_1}, but are also more prone to disseminating other types of toxic chat messages. 
These linguistic patterns could aid moderators in recognizing and addressing potential fanchuan users more effectively.

\section{\blue{Automated Identification of Fanchuan Users (RQ4)}}
\label{sec:rq4}

\blue{In this section, we explore whether a machine learning model can accurately identify fanchuan users.
This could be helpful in developing tools that more effectively moderate fanchuan behaviors during livestreams.}
The inherent complexity of fanchuan messages, characterized by their broad range of topics and the frequent use of context-specific abbreviations (as demonstrated in \S\ref{sec:RQ1}) presents challenges in creating a model focused on chat message level. Yet, our findings in \S\ref{sec:RQ3} reveal distinct behavioral patterns among fanchuan users. \blue{This leads us to develop a model that can identify and flag potential fanchuan \emph{users}, rather than chat messages.}

\subsection{\blue{Model Design}}
\label{subsec:rq4_1}

\blue{To identify fanchuan users, we aim to train a machine learning model} that generates a ranking for viewers of the livestream, according to their probability of later posting fanchuan messages. 
\blue{The model assigns a score from 0 to 1 to each viewers in the livestream, estimating their probability of later sending a fanchuan message. The ranking  of the top $n$ viewers most likely to be fanchuan users is then generated based on these scores.}

\pb{Feature Engineering.}
Drawing on insights from RQ3, we identify key characteristics that significantly differentiate fanchuan users from other users. 
Although our analysis in \S\ref{subsec:rq3_1} reveals that fanchuan users repeatedly engage in such behavior, we deliberately exclude ``previously sending fanchuan messages'' as a feature to avoid tautological reasoning. 
To encapsulate the spammer-like fanchuan users, as outlined in \S\ref{subsec:rq3_1} and \ref{subsec:rq3_2}, we incorporate the insight that these accounts tend to be newly created. This is operationalized by including the Bilibili ID as a feature, with the understanding that newer users have larger ID numbers. Additional features are derived from activity metrics discussed in  \S\ref{subsec:rq3_2}, as well as sentiment and toxicity metrics. A detailed summary of these features is provided in Table \ref{tab:rq4_feature}.

\pb{Model Training.}
We experiment with five machine learning algorithms: Linear Regression (LiR), Logistic Regression (LoR), Random Forest (RF), Histogram-Based Gradient Boosting (HGB), and K-nearest Neighbors (KNN). We apply these algorithms to training data from the Fanchuan User List and the Comparison User List, as outlined in \S\ref{subsec:dataset_construction}. 
Recall, the Comparison User List include the users who have sent at least one chat message within a 5-minute window preceding the fanchuan message. We do not include Fanchuan (SC) Users, as in such cases, the moderator can easily check every superchat given the low number of superchats.
To address class imbalance, we undersample the group of Comparison Users. 
We use 80:20 train-test split and then implement 5-fold cross-validation with grid search to optimize the hyperparameters for each model. The specific hyperparameters selected for each model are detailed in Table \ref{tab:rq4_parameter}.

\subsection{\blue{Model Evaluation}}
\label{subsec:rq4_2}

\pb{Evaluation Metric.}
To evaluate the effectiveness of the \blue{rankings generated by the model}, we utilize a ranking performance metric. This metric is determined by the rank position of users who actually send a fanchuan chat message.
More precisely, for each fanchuan chat message sent by a user (denoted as $F$) during a live streaming session, $S$, we take the list, $L_{comparison}$, of comparison users who have sent at least one chat message within the 5-minute period before the fanchuan message.
We then use the trained model to calculate the probability, $P_F^S$, that user $F$ will send a fanchuan message in session $S$. We also compute the probability, $P_C^S$, for each of the comparison users, $C$, in $L_{comparison}$, for the same session, $S$. After calculating these probabilities, we rank $P_F^S$ along with all $P_C^S$ values, and identify the rank position of $P_F^S$ in this list. The rank position of user $F$ defines the ranking performance metric.
A higher rank position signifies a more precise and effective ranking system, as it indicates that users who actually send a fanchuan chat message are positioned higher in the ranking.

\begin{figure}
    \centering
    \includegraphics[width=0.7\linewidth]{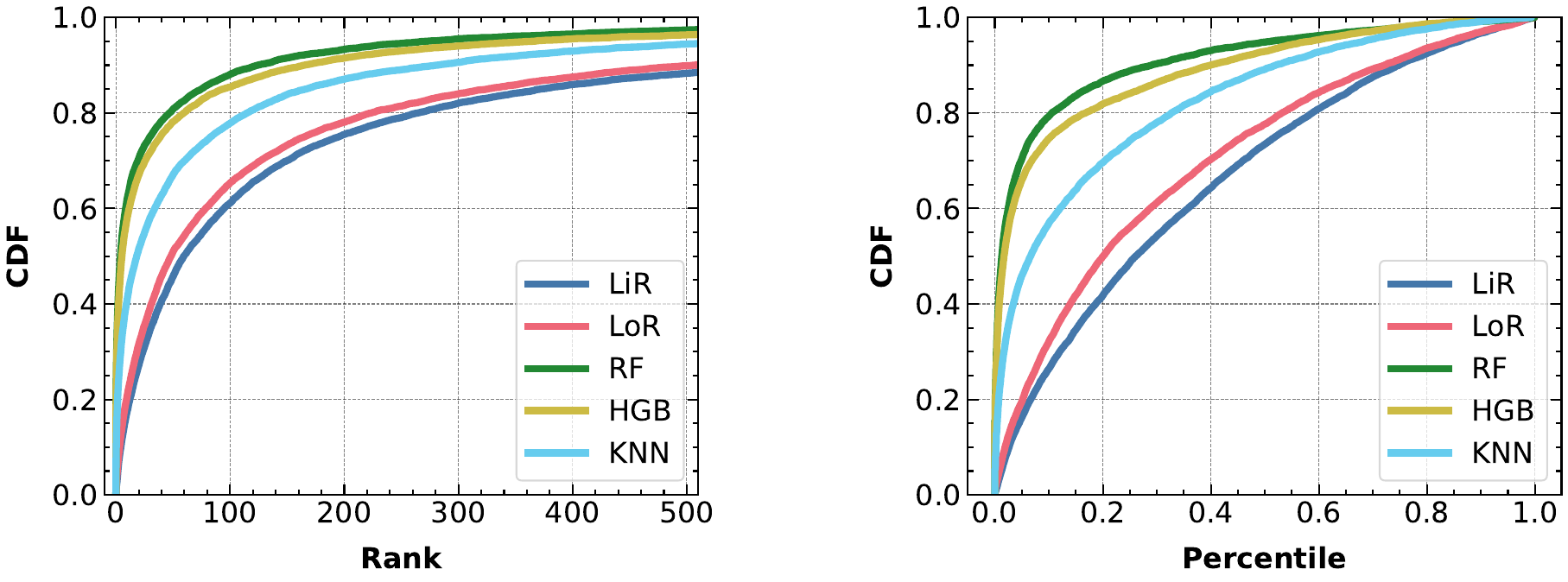}
    \caption{CDF of the rank and the percentile of the user who sends the fanchuan chat message among all viewers in the livestreaming session who send a chat message in the 5-minute period prior to the fanchuan message.}
    \label{fig:rq4_result}
\end{figure}

\pb{Results.}
We compute the ranking performance metric and obtain results for each fanchuan user in terms of both raw ranking position and percentile (\ie top $x$\% among the predicted viewers). The outcomes for the five models are depicted as CDF in Figure \ref{fig:rq4_result}. An effective prediction would ensure that all fanchuan users achieve a high rank. The results reveal that Random Forest attains the best performance, followed by Histogram-Based Gradient Boosting and K-Nearest Neighbors, while Linear Regression and Logistic Regression lag behind.

For the top-performing model (Random Forest), 34\% of the users who send a fanchuan message during the live streaming session are ranked in first place, while 53\% are among the top 5 positions and 81\% are among the top 50 positions. This demonstrates that, in most cases, the model can accurately identify viewers who are likely to send fanchuan messages within a limited group of users (\eg top 50).
Consequently, our findings confirm that \blue{our model} is capable of pinpointing a manageable number of, for instance, 50 viewers out of hundreds or thousands. 
\blue{This confirms the potential of using automated tools to support moderation for fanchuan behaviors, which we further discuss in next subsection \S\ref{subsec:rq4_3}.}

\subsection{\blue{Implications}}
\label{subsec:rq4_3}

\blue{The intended usage of our model to rank potential fanchuan users is for tool-supported human moderation.
Human moderation has traditionally been the cornerstone of managing large-scale interaction during livestreams. However, it presents its own set of challenges. Given the real-time nature of livestreams and the sheer volume of interactions, it can be overwhelming for moderators to keep up \cite{10.1145/3544549.3585704, 10.1145/3491102.3517628}.
Semi-automatic moderation tools are designed to alleviate the burden on human moderators and enhance the efficiency of moderation efforts \cite{10.1145/3491102.3501999, 10.1145/3359276}. These systems assist in identifying and potentially performing actions against certain behaviors. By analyzing patterns in chat messages and user interactions, machine learning algorithms can flag malicious behaviors for human verification by moderators.}

\blue{We believe that our model can be effective for supporting such human moderation. While previous work primarily focuses on the content (\eg chat messages) and flags harmful content, our model concentrates on the historical behavior of the user, making it possible to identify and flag potentially malicious users.
Currently, it is impractical for moderators to single out fanchuan chat messages from thousands of viewers. Moreover, fanchuan messages are inherently difficult to distinguish. With the assistance of our model, it is possible to develop a tool capable of identifying a small group of viewers (\eg 50 based on the results in  \S\ref{subsec:rq4_2}), who have a high likelihood of being, or becoming, fanchuan users. Consequently, their chat messages can be manually reviewed by a human moderator before being broadcast in the livestream, enabling effective moderation of fanchuan.}

\section{\minorblue{Broader Implications}}

\pb{\blue{Generalizability of the Study.}}
\blue{We examine a novel phenomenon that has emerged in a predominantly non-Western, non-English-speaking community, offering a valuable contribution to a less explored area of CSCW literature. We focus on Bilibili, the largest Twitch-like or YouTube-like livestreaming platform in China, boasting a daily active user of 97 million and a monthly active user of 315 million as of the first quarter of 2023 \cite{bilibili-2023q1}, making it the most important platform for studying such livestreaming in China. Thus, this study serves as an important case study. Furthermore, it is likely that our findings can be generalized to most other Chinese platforms, as they employ nearly identical interactive infrastructures: real-time chat systems, virtual gifting/donation mechanics, and moderation systems. Additionally, as demonstrated in \S\ref{sec:RQ1}, the phenomenon of fanchuan predominantly arises in platform-neutral content, such as online games and esports, which are widely available and popular across all mainstream platforms.}

\blue{That said, we note that, due to language and cultural differences, fanchuan may not be commonly observed on non-Chinese platforms like Twitch and YouTube, and our findings may not be directly applicable to them.
However, other new forms of antisocial behavior --— those that are covert (such as content leakage \cite{10.1145/3131365.3131385, 10.1145/3173574.3174241}), strategic (such as impersonation \cite{10.1145/3173574.3174241, 10.1145/3025453.3025875}), or collaborative (such as hate raids \cite{10.1145/3610191, 10.1145/3579609}) —-- are becoming increasingly common over all platforms in the world, posing new challenges compared to conventional forms \cite{9519435, 10.1145/3634737.3644998, 10.1145/3639362}. 
}
\blue{Fanchuan is a representative form of such a new type of attack, characterized by its covert and indirect nature, strategically designed for long-term damage. We have examined its negative impact, analyzed its characteristics, and explored potential moderation strategies.
Our analysis yields critical insights into the evolving landscape of online antisocial behaviors, exemplifying a paradigm shift from simple overt attacks to sophisticated tactics. The findings underscore the need for attention from researchers, platform designers, and policymakers to develop better approaches and techniques for moderation and platform governance. }

\pb{\blue{Evolution of Moderation Methods.}}
\blue{To effectively combat these evolving forms of antisocial behavior, it is crucial to develop new moderation strategies. Traditional content moderation methods, which often focus on detecting offensive content, are inadequate against these new forms of antisocial behavior \cite{10.1145/3359157, 10.1145/3359146}, where fanchuan is a good example. 
Additionally, using machine learning models to detect antisocial behavior content is reported to be challenging in the context of livestreaming chats  \cite{moon-etal-2023-analyzing, Jiang_Shen_Wen_Sha_Chu_Liu_Backes_Zhang_2024} and culture-specific Internet slang \cite{xiao-etal-2024-toxicloakcn}.
Previous research has shown that examining users' historical behavior can be helpful \cite{10.1145/3479554}. Indeed, we demonstrate that a machine learning model focusing on user profile history and activity, rather than solely on content, can enhance identification of potential fanchuan users. This could also aid in moderating other forms of covert and indirect attacks. However, implementing this approach may necessitate collaboration with platforms to provide relevant data and tools for moderators. The deployment of such approaches raises key ethical concerns, including issues related to privacy. We recommend that platforms thoughtfully design interfaces for trained moderators to access the necessary data for profiling and identifying potentially harmful users.}

\blue{Our work also shows that collaborative moderation approaches, which harness the collective efforts of community members, could offer a more robust defense against coordinated attacks. While previous methods have primarily focused on collaboration among moderators \cite{10.1145/3491102.3517628, 10.1145/3544548.3580982, 10.1145/3290605.3300390}, moderators alone may not always detect covert and indirect threats, whereas other users might be able to. This is how we identify fanchuan behavior and compile our dataset for this study. Additionally, many covert and indirect attacks, such as impersonation and misinformation, aim to create a false impression among users. Naturally, the impact of these attacks can be (partially) mitigated if a user can recognize and flag them for others, challenging this false impression.
However, without effective mechanisms, it is challenging for an normal viewer to deal with an attack, even if they identify it (\eg their chat message might be lost among thousands of others). Therefore, we recommend exploring new technologies to facilitate convenient collaborative moderation, leveraging the power of each community member. }

\blue{We believe that these technologies must evolve in tandem with the tactics they aim to counteract, ensuring they remain effective in preserving the integrity and safety of digital spaces. Additionally, the role of platforms and stakeholders in governance and safety is crucial for addressing these evolving challenges. Platforms must adopt proactive, up-to-date measures to establish and enforce comprehensive, evolving moderation policies that protect users from new forms of attacks.}

\section{Conclusion}

\pb{Summary.}
This paper has examined fanchuan attacks, a unique and emerging type of malicious behavior typically found in livestreaming chat. We conduct the first empirical study on fanchuan behavior, focusing on Bilibili. Our dataset comprises 2.7 million livestreaming sessions on Bilibili, featuring 3.6 billion chat messages, where We identify 130,000 instances of fanchuan behavior across 37,400 livestreaming sessions. Our research provides valuable insights into fanchuan behavior and its perpetrators, which we leverage to \blue{show that a machine learning model can effectively identify a small groups of potential fanchuan users.}

\pb{Future Work.}
Our study focuses on livestreaming chats, yet we note that fanchuan attacks can also appear on other social media platforms such as micro-blogs and forums. In Bilibili livestreaming, there is a large volume of real-time chat messages, and the user's ID is not directly visible. However, it is arguably easier on other social media for users to identify fanchuan behavior by checking the user's profile and activity history. In such cases, fanchuan users may employ more complex methods, like using accounts specifically for fanchuan activities. Therefore, in our future work, we plan to expand our research on fanchuan behavior to other platforms.
While our research reveals several characteristics of fanchuan users, the underlying motives remain unclear. Thus, in order to gain a deeper understanding, we also intend to use qualitative methods, including questionnaires and interviews, in our future work. 
\blue{In Section \S\ref{sec:rq2_2}, we discuss the chat sentiment and chat toxicity changes before and after the fanchuan behavior. For future work, we would like to manually annotate more specific times of the fanchuan behavior to break down the fanchuan behavior more in terms of time span, not limited to pre-fanchuan and post-fanchuan, which will allow us to understand more about the sentiment and toxic effects of fanchuan behavior.}
\blue{In Section \S\ref{subsec:rq4_3}, we discuss the intended use of the machine learning model for identifying potential fanchuan users, with the goal of developing a moderation tool to assist moderators in managing fanchuan more effectively. For future work, we would also like design, implement, and evaluate this moderation tool.}

\subsection*{Acknowledgments}
This work was supported in part by the Guangzhou Science and Technology Bureau (2024A03J0684), Guangdong provincial project 2023QN10X048, the Guangzhou Municipal Key Laboratory on Future Networked Systems (2024A03J0623), the Guangdong Provincial Key Lab of Integrated Communication, Sensing and Computation for Ubiquitous Internet of Things (No.2023B1212010007), the Guangzhou Municipal Science and Technology Project (2023A03J0011), Guangdong provincial project (2023ZT10X009), and the 111 Center (No. D25008).

\bibliographystyle{ACM-Reference-Format}
\bibliography{sample-base}

\appendix
\newpage
\section{Appendix}

\subsection{Data Description}
\label{subsec:appendix_data_description}

\begin{table}[h!]
\centering \small
\begin{tabular}{llL{28em}}
\hline\hline
{}           & \textbf{Type}    & \textbf{Description} \\
\textbf{Field}           & {}    & {} \\ 
\hline 
\texttt{uId}             & Integer          & Unique identifier for the streamer.                                                   \\ 
\texttt{uName}           & String           & Name of the streamer.                                \\ 
\texttt{liveId}          & String           & Unique identifier for the live session.                                           \\ 
\texttt{parentArea}      & String           & The parent category or area of the live session.                                  \\ 
\texttt{area}            & String           & The specific area or category of the live session.                                \\ 
\texttt{coverUrl}        & String           & URL of the cover image for the live session.                                      \\ 
\texttt{startDate}       & Integer (Epoch)  & Start time of the live session, represented as a Unix timestamp.                  \\ 
\texttt{stopDate}        & Integer (Epoch)  & Stop time of the live session, represented as a Unix timestamp.                   \\ 
\texttt{title}           & String           & Title of the live session.                                                        \\ \hline\hline
\end{tabular}
\caption{Description of data fields for live session}
\label{table:data_description}
\end{table}

\begin{table}[h!]
\centering \small
\begin{tabular}{llL{28em}}
\hline\hline
{}          & \textbf{Type}             & \textbf{Description}  \\
\textbf{Field}           & {}             & {}                                                       \\ \hline
\texttt{uId}             & Integer                   & Unique identifier for the user who sent the interaction.                             \\ 
\texttt{uName}           & String                    & Name of the user who sent the interaction.                                           \\ 
\texttt{type}            & Integer                   & Type of the interaction.                                                             \\ 
\texttt{sendDate}        & Integer (Epoch)           & Time when the interaction was sent, represented as a Unix timestamp.                 \\
\texttt{message}         & String                    & Content of the interaction message.                                                  \\ 
\texttt{price}           & Integer                   & Price associated with the interaction, if any.                                       \\ 
\texttt{count}           & Integer                   & Count associated with the interaction, if any.                                       \\ \hline\hline
\end{tabular}
\caption{Description of data fields for viewer interaction}
\label{table:danmakus_description}
\end{table}

\subsection{Prompts to Process Search Result}
\label{subsec:appendix_prompts_filt}

\promptbox{
\footnotesize
\\
\systemmessage~
You are a helpful assistant designed to output JSON. \\
\usermessage~
\{
    "task": "Classify given live chat messages. Output the results in JSON list." 
    "description": "The term \zh{'反串'} refers to pretending to support or like something in an exaggerated or overly enthusiastic way, while actually disliking or criticizing it. The person engaging in this behavior is called a \zh{'反串'}. To ask someone to stop this behavior, one can say \zh{'反串'}.", 
    "classification": \{ 
        "0": "Not related to \zh{'反串'}", 
        "1": "Message is related to \zh{'反串'}, and potentially indicates that someone is \zh{'反串'}", 
        "2": "Message is related to \zh{'反串'}, but clearly not indicating that someone is \zh{'反串'}"
    \},
    "input\_format": ["text\_1", "text\_2", "..."],
    "output\_format": ["label\_1", "label\_2", "..."],
    "input": ["the input list of chat messages to be processed"]
\}
}

\subsection{Prompts to Extract Nouns}
\label{subsec:appendix_prompts_nouns}

\promptbox{
\footnotesize
\\
\systemmessage~
You are a helpful assistant designed to output JSON. \\
\usermessage~
\{
    "task": "Given a text input, extract all name entities and nouns into a JSON list."
    "input\_format": ["text\_1", "text\_2", "..."],
    "output\_format": ["task\_result\_of\_text\_1", "task\_result\_of\_text\_2", "..."],
    "input": ["the input list of chat messages to be processed"]
\}
}

\subsection{Manual Review of Fanchuan Chat Messages}
\label{subsec:appendix_manual}
For each Fanchuan chat message, the authors manually compare it with other chat messages sent 0 to 15 seconds prior. The authors also consider the title and area of the livestream. The authors then determine if: \one the Fanchuan message is related to previous chat messages, \two it is only related to the title or area of the livestream, or \three it is unrelated to either.

\subsection{Additional Figures for Section \ref{sec:RQ3}}
\begin{figure}[h!]
    \centering
    \includegraphics[width=\linewidth]{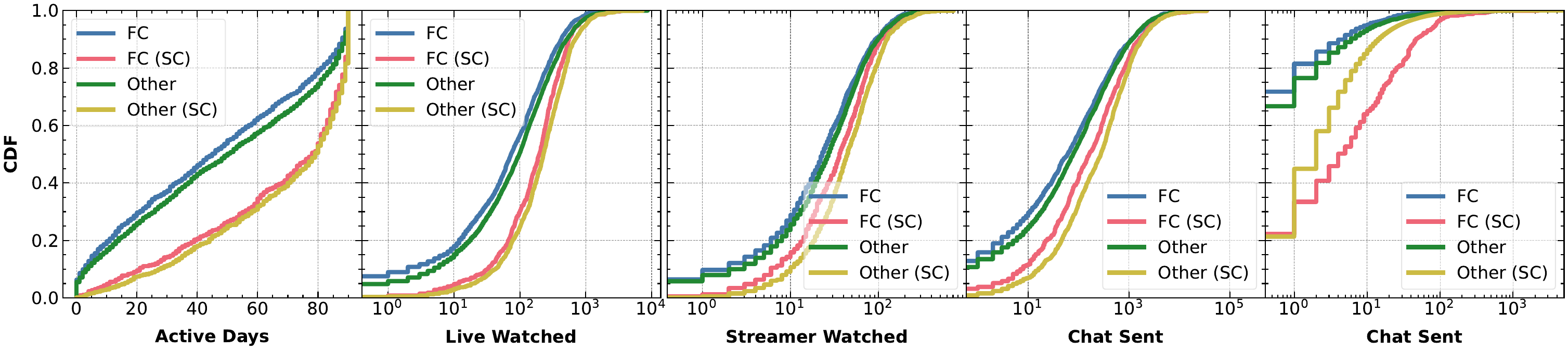}
    \caption{CDF of the activity metrics for Fanchuan Users and Comparison Users, focusing solely on livestreaming sessions in the same area where Fanchuan behavior occurs, during the measurement period.}
    \label{fig:rq3_activity_area}
\end{figure}




\subsection{Features}
\label{subsec:appendix_features}

\begin{table}[h!]
    \centering
    \small
    \begin{tabular}{llL{20em}}
    \hline\hline
         {}  & \textbf{Measured} & \textbf{Description} \\
         \textbf{Feature} & {} & {} \\
    \hline
        Active Days  & Overall$^1$ \& Other Area$^2$ & Number of active days \\
        Live Watched  & Overall \& Other Area &  Number of livestreaming session watched \\
        Streamer Watched  & Overall \& Other Area & Number of unique streamer watched \\
        Chat Sent  & Overall \& Other Area &  Number of chat messages sent \\
        Gift \& SC Sent  & Overall \& Other Area &  Number of gift and superchat sent \\
        \$ Gift \& SC Sent  & Overall \& Other Area &  Monetary value of gift and superchat sent \\
        Toxicity Rate  & Overall & Proportion of toxic chat messages out of all chat messages sent \\
        Positive Sentiment Rate  & Overall & Proportion of chat messages of positive sentiment out of all chat messages sent \\ 
        Negative Sentiment Rate  & Overall & Proportion of chat messages of negative sentiment out of all chat messages sent \\
        Neutral Sentiment Rate & Overall & Proportion of chat messages of neutral sentiment out of all chat messages sent \\
        ID Number  & - & Bilibili ID number in base 10 logarithm \\
    \hline\hline
    \end{tabular}
    \\
    \begin{flushleft}
        1. Result measured across the platform in the 90-day period before the current livestreaming session. \\
        2. Result measured for livestreaming sessions in different area from the current livestreaming session, in the 90-day period before the current livestreaming session.
    \end{flushleft}
    \caption{Features used for machine learning models.}
    \label{tab:rq4_feature}
\end{table}

\newpage
\subsection{Hyperparameters}
\label{subsec:appendix_parameters}

\begin{table}[h!] 
    \centering 
    \small
    \begin{tabular}{lL{20em}} 
    \hline\hline
    \textbf{Algorithm} & \textbf{Parameters} \\ 
    \hline 
        Linear Regression & - \\ 
        Loistic Regression & penalty=L2, C=1.0 \\ 
        Random Forest & n\_estimators=100, max\_depth=16 \\ 
        Histogram-Based Gradient Boosting & learning\_rate=0.1, max\_iter=100, max\_depth=None \\ 
        K-Nearest Neighbors & n\_neighbors=100, leaf\_size=30, p=2, metric=minkowski \\ 
    \hline\hline
    \end{tabular} 
    \caption{Hyperparameters Used for Each Model.} 
    \label{tab:rq4_parameter} 
\end{table}

\end{document}